\documentclass{article}
\usepackage{arxiv}
\renewcommand{\undertitle}{}
\renewcommand{\headeright}{}
\date{}
\usepackage{microtype}
\usepackage[normalem]{ulem}
\usepackage{xspace}
\usepackage{tikz}
\usetikzlibrary{pgfplots.groupplots}

\usepackage{pifont}
\usepackage{booktabs}

\usepackage{pgfplots}
\usepackage{pgfplotstable}
\usepackage{amsmath,amsfonts,amsthm,enumitem}
\usepackage[linesnumbered,ruled]{algorithm2e}
\usepackage{graphicx}
\usepackage{url}

\usepackage{textcomp}
\usepackage{xcolor}
\usepackage[labelfont=bf]{caption}
\usepackage{subcaption}

\usepackage[T1]{fontenc}
\usepackage[utf8]{inputenc}

\usepackage{colortbl}
\usepackage[most]{tcolorbox}
\usepackage{listings}
\usepackage{makecell}

\usetikzlibrary{patterns,arrows}
\usepackage{multicol}
\usepackage{multirow}
\usepackage{wrapfig}

\usepackage{tabularx,array}

\definecolor{accA}{HTML}{2F5D8C}
\definecolor{accB}{HTML}{B5562B}
\definecolor{accC}{HTML}{2E7D5B}
\definecolor{ink}{HTML}{2B2B2B}
\definecolor{meta}{HTML}{8A8A8A}

\newcommand{\gro}{{\color{meta}\boldmath$\bigr)$\kern0.5pt?}}

\newtcolorbox{panel}[4][]{%
  enhanced, breakable=false,
  colback=white, colframe=#2!55, boxrule=0.6pt, arc=2pt,
  left=6pt, right=6pt, top=2pt, bottom=3pt,
  before skip=0pt, after skip=9pt,
  attach boxed title to top left={xshift=6pt,yshift=-\tcboxedtitleheight/2},
  boxed title style={colback=#2, colframe=#2, arc=1.5pt, boxrule=0pt,
                     left=3pt,right=5pt,top=1pt,bottom=1pt},
  fonttitle=\sffamily\bfseries\small\color{white},
  title={\colorbox{white}{\color{#2}\makebox[0.8em]{#3}}\hspace{3pt}#4},
  before upper={\vspace{3pt}}, #1
}

\tcbset{flowto/.style={overlay={%
  \draw[-{latex},line width=0.8pt,color=meta] (frame.south)--++(0,-9pt);
  \node[right,font=\sffamily\scriptsize\itshape,color=meta,inner sep=1pt]
       at ([xshift=3pt,yshift=-4.5pt]frame.south){#1};}}}

\usetikzlibrary{trees}

\usepackage[]{mdframed}

\setlist[itemize]{
    topsep=2pt,
    itemsep=1pt,
    parsep=0pt,
    partopsep=0pt
}

 \definecolor{catIV}{HTML}{4E79A7}
 \definecolor{catSC}{HTML}{F28E2B}
 \definecolor{catSI}{HTML}{59A14F}
 \definecolor{catDT}{HTML}{B07AA1}
 \definecolor{catPI}{HTML}{E15759}

\definecolor{dkgreen}{rgb}{0,0.6,0}
\definecolor{gray}{rgb}{0.5,0.5,0.5}
\definecolor{mauve}{rgb}{0.58,0,0.82}
\definecolor{orangep}{rgb}{0.71, 0.43, 0.89}
\definecolor{orp}{rgb}{1, 0.7, 0.278}

\definecolor{darkBlue}{rgb}{0.000000,0.000000,0.545098}
\definecolor{darkGreen}{rgb}{0.000000,0.392157,0.000000}
\definecolor{DarkGray}{gray}{0.4}
\definecolor{javared}{rgb}{0.6,0,0}
\definecolor{javagreen}{rgb}{0.25,0.5,0.35}
\definecolor{javapurple}{rgb}{0.5,0,0.35}
\definecolor{javadocblue}{rgb}{0.25,0.35,0.75}
\definecolor{lightgray}{gray}{0.95}
\definecolor{shadecolor}{RGB}{150,150,150}
\definecolor{blueA}{RGB}{204,229,255}
\definecolor{redA}{RGB}{112,0, 0}
\lstdefinestyle{MyCSmallStyle} {
  language=C++,
  frame=none,
  xleftmargin=15pt,
  stepnumber=1,
  numbers=left,
  numbersep=5pt,
  numberstyle=\tiny\color[black]{0.177},
  belowcaptionskip=\bigskipamount,
  captionpos=b,
  escapeinside={*'}{'*},
  tabsize=5,
  emphstyle={\bf},
  escapechar=!,
  basicstyle=\scriptsize\ttfamily,
  keywordstyle=\color{javapurple}\bfseries,
  stringstyle=\color{javared},
  commentstyle=\color{javagreen},
  morecomment=[s][\color{javadocblue}]{/**}{*/},
  showspaces=false,
  columns=flexible,
  showstringspaces=false,
  morecomment=[l]{//},
  tabsize=2,
  breaklines=true,
  moredelim=[is][\underbar]{^}{^}
}
\lstdefinestyle{MyJavaSmallStyle} {
  language=Java,
  frame=none,
  xleftmargin=15pt,
  stepnumber=1,
  numbers=left,
  numbersep=5pt,
  numberstyle=\color{DarkGray},
  belowcaptionskip=\bigskipamount,
  captionpos=b,
  escapeinside={*'}{'*},
  tabsize=5,
  emphstyle={\bf},
  basicstyle=\scriptsize\ttfamily,
  keywordstyle=\color{javapurple}\bfseries,
  stringstyle=\color{javared},
  commentstyle=\color{javagreen},
  morecomment=[s][\color{javadocblue}]{/**}{*/},
  showspaces=false,
  columns=flexible,
  showstringspaces=false,
  morecomment=[l]{//},
  tabsize=2,
  breaklines=true,
  moredelim=[is][\underbar]{^}{^}
}

\lstdefinelanguage{Scala}{
  keywords={typeof, new, true, false, catch,def,val, function, return, null, catch, switch, var, if, in, while, do, else, case, break, assert, static, void ,declare, const, for, define,fun, ite,class, not, check,sat,String, Int, ArrayList},
  keywordstyle=\color{blue}\bfseries,
  ndkeywords={ export,extends, boolean, throw, implements, import, this, abstract,reduceByKey, reduce, filter, map, reduceByKey, join, Join1, public },
  ndkeywordstyle=\color{mauve}\bfseries,
  otherkeywords={+, =>,<=, ==, >,< , ||},
  identifierstyle=\color{black},
  sensitive=false,
  comment=[l]{//},
  morecomment=[s]{/*}{*/},
  commentstyle=\color{purple}\ttfamily,
  stringstyle=\color{red}\ttfamily,
  morestring=[b]',
  morestring=[b]"
}

\definecolor{assertred}{RGB}{180,30,30}

\lstdefinestyle{pythonbase}{
  language=Python,
  basicstyle=\ttfamily\small,
  columns=fullflexible,
  keepspaces=true,
  showstringspaces=false,
  frame=single,
  breaklines=true,
  linewidth=\linewidth,
  keywordstyle={},
}

\lstdefinestyle{pythonassert}{
  style=pythonbase,
  moredelim=**[is][\color{assertred}\bfseries]{@}{@},
}

\newcommand{\MyPara}[1]{\vspace{.2em}\noindent\textbf{\textit{#1}}~}
\newcommand{\codefont}[1]{{\texttt{#1}}}

\newcommand{\ie}{\emph{i.e.,}\xspace}

\newcommand{\tool}{{\small{\textsc{XCheck}}}\xspace}
\newcommand{\toolnoprioritization}{{\small{\sc{\tool-Random}}}\xspace}

\newcommand{\neuri}{{\small{\textsc{NeuRi}}}\xspace}
\newcommand{\gencog}{{\small{\textsc{GenCoG}}}\xspace}
\newcommand{\deeprel}{{\small{\textsc{DeepREL}}}\xspace}
\newcommand{\hirgen}{{\small{\textsc{HirGen}}}\xspace}

\usepackage[square,comma,numbers,sort]{natbib}
\usepackage[hidelinks]{hyperref}

\title{Revealing Compiler-Platform Interaction Issues in Deep Learning Pipelines through Cross-Layer Constraints}

\author{
\textbf{
Yuxin Qiu\thanks{Corresponding email: \href{mailto:yuxin.qiu@email.ucr.edu}{yuxin.qiu@email.ucr.edu}.}\textsuperscript{1},
Jiyuan Wang\textsuperscript{2},
Geyang Xu\textsuperscript{1},
Ben Limpanukorn\textsuperscript{3},
Miryung Kim\textsuperscript{3},
Qian Zhang\textsuperscript{1}
}\\
{
\textsuperscript{1}UC Riverside
\quad
\textsuperscript{2}Tulane University
\quad
\textsuperscript{3}UCLA
}
}

\begin{document}

\maketitle

\begin{abstract}

The growing deployment of artificial intelligence (AI) increases the need for reliable deep learning (DL) compilers, such as TVM and ONNX-MLIR. These compilers lower high-level DL models through multiple intermediate representations and specialize them for different hardware platforms. Their correctness therefore depends on constraints from multiple layers of the compilation stack. Such constraints include model validity, compiler transformation requirements, and target-platform limitations.

Existing DL compiler testing techniques have made substantial progress in generating valid and diverse models and in exercising compiler optimizations. However, they represent only part of the constraints that are required to determine whether a model can be compiled and deployed correctly on a target platform. As a result, compiler--platform interaction issues remain hidden: generated tests may be rejected before reaching hardware-dependent compilation logic, and differences in what the compiler computes along the way do not show in the compilation result.

In this work, we propose \tool, a DL compiler testing framework that reveals such issues through cross-layer constraints. \tool deterministically extracts constraints from model and compiler documentation, compiler implementation, and hardware specifications, and it checks that each constraint can be violated on its own and that each guard derived from the constraints is consistent with the transformation it monitors. It then uses the constraints in two ways. First, \tool synthesizes models that satisfy all applicable constraints or violate a targeted one, and it prioritizes constraints by the compilation stage that their previous tests reached and by how often they have been selected. Second, \tool inserts guards into the compilers that record how intermediate compilation states relate to quantities of the compiler and the platform, which distinguishes compilations that coverage and compilation outcomes treat as the same.

We evaluate \tool on three DL compilers against four baselines.
\tool extracts 987 cross-layer constraints, whereas no baseline encodes a constraint that involves the platform.
It detects 36 issues, and the baselines miss 72.22\% of them.
Of these issues, 21 are compiler--platform interaction issues, and the baselines miss 85.71\% of them.
The developers have since fixed eight issues.
Constraint prioritization raises the share of tests that reach code generation by up to 8.50 percentage points over selecting constraints uniformly at random, and with its guards, \tool observes 269 distinct behaviors of intermediate compilation states across the three compilers.

\end{abstract}

\section{Introduction}
\label{sec:introdction}

Deep learning (DL) compilers translate neural network models into executable code for heterogeneous hardware through a series of intermediate representation (IR) transformations. Compilers such as TVM~\cite{tvm}, TensorRT~\cite{tensorrt}, ONNX-MLIR~\cite{onnx-mlir}, and GeneSys~\cite{genesys} progressively lower model computations while optimizing data layouts, memory allocation, and execution schedules. These transformations must satisfy constraints imposed by the model, the compiler, and the target hardware. For example, GeneSys lowers an ONNX model~\cite{onnx} into {\it codelets}, an IR that describes how each operator runs on the accelerator. It then maps the codelets onto hardware with fixed memory sizes, bandwidth, and array dimensions. Correct compilation therefore requires more than a valid model: transformations throughout the compilation pipeline must satisfy interdependent constraints across model semantics, compiler operations, and hardware resources.
A compilation that breaks one of these constraints exposes a {\it compiler--platform interaction issue}, which can surface as a crash, a valid model that the compiler wrongly rejects, or an invalid model that it wrongly accepts.

\begin{figure}[t]
  \centering
  \includegraphics{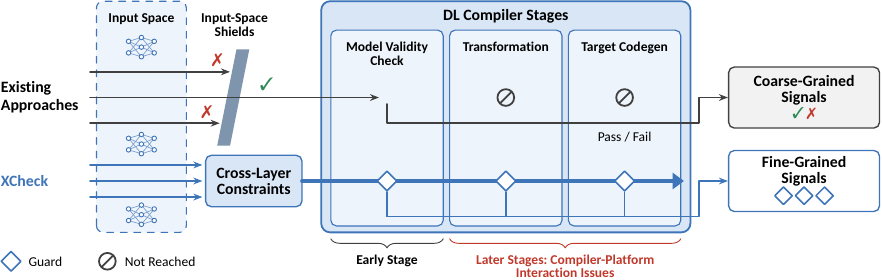}
  \caption{Existing approaches and \tool in DL compiler testing. Existing approaches~\cite{nnsmith,neuri,gencog,hirgen,deeprel,modelmeta} encode mainly model-level constraints and check only the compilation outcome. Their tests can stop at the model validity check, and the outcome does not show the intermediate compilation states. \tool uses cross-layer constraints to guide its tests to the later compilation stages, and its guards record the intermediate states at each stage.}

  \label{fig:position}
\end{figure}

\MyPara{Current Practices and Limitations.}
Existing DL compiler testing techniques primarily generate models or compiler IRs and check their compilation outcomes~\cite{nnsmith,neuri,gencog,hirgen,mlirsmith,polyjuice}.
Constraint-based generators such as NNSmith~\cite{nnsmith}, NeuRI~\cite{neuri}, and GenCoG~\cite{gencog} construct valid models using operator constraints. Transformation-based approaches such as HirGen~\cite{hirgen} and PolyJuice~\cite{polyjuice} derive tests through IR or model transformations. Other techniques use compiler implementation details to target specific compiler logic~\cite{whitefox,oatest}. These techniques explore diverse model structures and optimizations.
As shown in {Figure~\ref{fig:position}}, while these approaches expand the space of tested models and transformations, they do not systematically capture the cross-layer dependencies that govern which compilation stages a test reaches and what intermediate results the compiler computes. This gap creates three fundamental challenges.

First, model validity does not guarantee hardware feasibility. For example, GeneSys crashed during FPGA memory allocation for a model with tensor shape $(1,16,10048934087,22145136773)$. The model passed the model validity check, but its tensor storage exceeded the available DRAM capacity. A constraint such as \codefont{TensorSize <= DeviceMemory} relates the tensor dimensions and data representation to the resource limit of the platform. Operator shape compatibility alone does not capture this requirement.

Second, constraint selection affects which compiler stages a test reaches. A shape mismatch may cause rejection during the model validity check. A resource constraint may become relevant during memory planning or backend lowering. In our GeneSys experiments, 61.60\% of the tests generated by NeuRI stop at the model validity check because they contain an operator that GeneSys does not support. Preserving the applicable model constraints while violating a resource constraint can direct testing toward later compiler stages. Compilation feedback can then guide subsequent constraint selection.

Third, reaching a pass does not show what the pass computes. Consider a padding transformation for a systolic-array target, which derives a padding factor from the target alignment and pads each dimension to a multiple of that factor. Two tests can execute the same padding code, while one of them pads a dimension and the other needs no padding because its dimension is already aligned. Coverage and compilation outcomes record the two executions in the same way, so they cannot distinguish how the compiler handles the two tests.

These challenges reveal two complementary roles of cross-layer constraints in DL compiler testing. They govern which model configurations can reach specific compilation stages, and they relate the intermediate values of each transformation to the parameters of the platform. Effective testing must therefore use constraints not only to construct valid models, but also to direct compilation toward later stages and to observe intermediate compilation states at a finer granularity than coverage and compilation outcomes.

Constraints relate quantities from three compilation layers: the model, such as graph connectivity, tensor shapes, data types, and operator attributes; the compiler, such as transformation preconditions and intermediate results; and the platform, such as supported representations, hardware alignment, and resource limits. {\it Cross-layer constraints} relate quantities across these layers, providing a common foundation for test generation and behavioral monitoring.

\begin{figure}[t]
  \centering
  \includegraphics{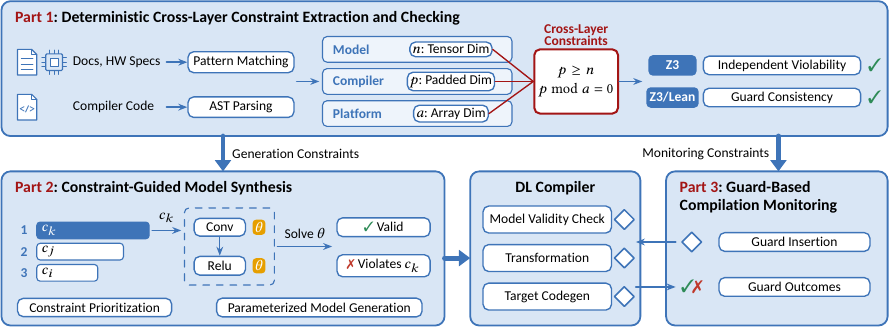}
  \caption{Overview of \tool. Cross-layer constraints are the foundation of \tool's model synthesis and compilation monitoring.}
  \label{fig:overview}
\end{figure}

\MyPara{\large \tool.}
We propose \tool, a DL compiler testing framework that unifies constraint-guided compilation exploration and guard-based behavioral monitoring through cross-layer constraints. The approach has three components.

\noindent\textit{Part 1. Deterministic Cross-Layer Constraint Extraction and Checking.}
\tool\ {\it deterministically} extracts operator constraints from the ONNX documentation and platform constraints from hardware specifications with rule-based parsing, and it analyzes the abstract syntax tree (AST) of the compiler source to refine these constraints and extract the constraints of compiler transformations, without AI-based inference.
\tool uses these constraints in two ways. {\it Generation constraints} involve only quantities known before compilation and guide model synthesis, whereas {\it monitoring constraints} involve values that the compiler computes and become guards in the compiler.
Because a correct rule can still yield a generation constraint that other constraints imply, or a monitoring constraint that is stricter than its transformation, \tool checks that each generation constraint can be violated while all others hold, and that each monitoring constraint is consistent with its transformation, with Z3 or Lean for arithmetic constraints and with manual review for the others.

\noindent\textit{Part 2. Constraint-Guided Model Synthesis.}
Compiler--platform interaction issues arise in later compilation stages, so a test can only reveal them if it gets through the earlier stages.
\tool therefore uses the extracted constraints to steer its tests toward later stages.
In each iteration, it prioritizes one constraint and synthesizes an ONNX model around the operator of this constraint that either satisfies all applicable constraints or violates the prioritized constraint.
Which constraint \tool prioritizes decides how far the test goes.
For example, violating a tensor-shape constraint often stops the test at the model validity check, whereas a test that stresses a platform constraint can pass the frontend and reach memory planning or target-specific lowering.
\tool therefore records the number of passes that the tests of each constraint complete and favors the constraints whose tests reach later stages.
Among constraints with equal shallowness, it prefers the ones it has prioritized less often, so that its tests keep exploring new compilation behaviors instead of repeating the same ones.

\noindent\textit{Part 3. Guard-Based Compilation Monitoring.}
Even when a test reaches a later stage, coverage and the compilation outcome describe its compilation only coarsely: they record which code runs and whether the compilation succeeds, but not what the compiler computes.
\tool therefore uses its monitoring constraints as finer-grained behavioral signals.
It inserts them as guards at the padding, tiling, loop-unrolling, and operator-fusion sites of the compiler, and each guard records the intermediate values at its site and whether its condition holds on them.
For example, at the padding site of GeneSys, a guard relates the padded dimension to the SIMD width of 16, so it separates a test that pads a dimension of 1 to 16 from a test whose dimension of 64 needs no padding, although both run the same pass and compile to the end.
For each compilation, \tool records which guards it evaluates, on which codelet, operator, or fused block, and with what outcome. These records distinguish compilations that coverage and compilation outcomes treat as the same.

\MyPara{Results and Contributions.}
We evaluate \tool on three DL compilers, GeneSys, ONNX-MLIR, and TVM, against four baselines, NeuRI~\cite{neuri}, DeepREL~\cite{deeprel}, GenCoG~\cite{gencog}, and HirGen~\cite{hirgen}, in four-hour campaigns.
\tool extracts 987 cross-layer constraints, which relate quantities from two or more of the model, compiler, and platform layers, whereas no baseline encodes a constraint that involves the platform.
It detects 36 issues, and the baselines miss 72.22\% of them.
Of these issues, 21 are compiler--platform interaction issues, and the baselines miss 85.71\% of them.
The developers have since fixed eight issues.
Compared with selecting constraints uniformly at random, constraint prioritization raises the share of tests that reach code generation by 8.15, 8.50, and 1.84 percentage points on TVM, ONNX-MLIR, and GeneSys, respectively.
With its 30 guards, \tool observes 269 distinct behaviors of intermediate compilation states across the three compilers, which the compilation result alone does not reveal.
This work makes the following contributions:
\begin{itemize}[leftmargin=*]
    \item We deterministically extract 987 cross-layer constraints for three DL compilers from documentation, compiler source code, and hardware specifications. We also check that each generation constraint can be violated alone and that each monitoring constraint is consistent with its transformation.
    \item We use the generation constraints to synthesize valid and targeted-violation models and prioritize the constraints whose tests reach later compilation stages. These models expose 36 issues in three DL compilers.
    \item We insert the monitoring constraints into the compilers as guards, which record intermediate compilation states that coverage and compilation outcomes do not distinguish.
\end{itemize}

\section{Background}
\label{sec:background}

\subsection{DL Models and DL Compilers}
DL models are directed graphs of tensor operators that process input tensors to generate output tensors. DL compilers translate DL models into executable code for {\it target hardware platforms}. While there exist various DL compilers, they typically follow a common compilation flow. For example, compilers that take ONNX-formatted models~\cite{onnx}, including TVM~\cite{tvm}, ONNX-MLIR~\cite{onnx-mlir}, and GeneSys~\cite{genesys}, broadly include three stages:
(1) {\it Model Validity Check:} This stage checks the correctness of ONNX models by verifying the compatibility of tensors and operators. For example, TVM first validates the basic structure of the model. If the model fails validation, the compilation is terminated; otherwise, TVM converts the model into a graph-level IR, such as \codefont{Relay} or \codefont{Relax}, for further optimization;
(2) {\it Multi-Layer IR Optimization:} This stage transforms the model into various IRs by applying high- and low-level transformations. High-level transformations restructure computation graphs, which are often hardware-agnostic, while low-level transformations optimize execution for specific hardware targets~\cite{shen2021study};
and (3) {\it Target Code Generation:} This stage converts the optimized IR into hardware-specific code for execution on target backends, such as CPUs, GPUs, FPGAs, and other specialized accelerators.

\subsection{Target Hardware Platforms}
\begin{table}[t]
\centering
\small
\caption{Common IR transformations in DL compilers, the value that the compiler chooses in each transformation, and the platform parameters that this choice depends on for systolic arrays and SIMD units.}
\label{tab:hardware}
\begin{tabularx}{\linewidth}{@{}l l l >{\raggedright\arraybackslash}X >{\raggedright\arraybackslash}X@{}}
\toprule
\multirow{2}{*}[-2pt]{\textbf{Transformation}} & \multirow{2}{*}[-2pt]{\textbf{IR Level}} & \multirow{2}{*}[-2pt]{\textbf{Chosen by Compiler}} & \multicolumn{2}{c}{\textbf{Platform Parameters}} \\
\cmidrule(l){4-5}
& & & \textbf{Systolic Array} & \textbf{SIMD} \\
\midrule
Operator fusion & High-level (graph) & Operators to fuse & \multicolumn{2}{c}{Largely hardware-agnostic} \\
\midrule
Tiling & Low-level (loop) & Tile sizes & Array dimensions; buffer capacity & Vector width; buffer or cache capacity \\
\midrule
Padding & Low-level (loop) & Padded sizes & Array dimensions & SIMD width; memory alignment \\
\midrule
Loop unrolling & Low-level (loop) & Unroll factors & Array dimensions (spatial unrolling) & Vector width; number of registers \\
\bottomrule
\end{tabularx}
\end{table}

IR optimizations in DL compilers are often coupled with target hardware, which offers different execution paradigms~\cite{fowers2018aconfigurable,xu2023asurvey,chen2016eyeriss}. {\it Single Instruction, Multiple Data (SIMD)}~\cite{fowers2018aconfigurable,liu2019optimizing} applies the same operation on multiple data points simultaneously. {\it Systolic arrays}~\cite{xu2023asurvey,das2020asystolic,genc2021gemmini,kung2019packing,sangkug2020flexsa} optimize matrix operations by streaming data through processing elements in a synchronized manner, which requires careful alignment between data and hardware.

Table~\ref{tab:hardware} lists IR transformations that DL compilers commonly apply and the platform parameters that their decisions depend on.
High-level transformations, such as {\it operator fusion} to reduce memory accesses, restructure the computation graph and are largely hardware-agnostic.
In contrast, low-level transformations choose their parameters from the platform.
For example, {\it tiling} sizes each tile to fit an on-chip buffer, and {\it padding} extends a tensor dimension to a multiple of the systolic-array dimension or the SIMD width, so the same model is tiled and padded differently on a systolic array and on a SIMD unit.

\begin{table*}[t]
\centering
\caption{Examples of cross-layer constraints. M: model; C: compiler, a restriction it imposes or a value it computes; P: platform.}
\label{tab:constr-category}
\begingroup
\scriptsize
\setlength{\tabcolsep}{3pt}
\renewcommand{\arraystretch}{1.15}
\begin{tabular}{@{}p{0.11\linewidth}p{0.18\linewidth}>{\raggedright\arraybackslash}p{0.56\linewidth}p{0.1\linewidth}@{}}
\toprule
\textbf{Use} & \textbf{Category} & \textbf{Example of Constraint Representation} & \makecell[l]{\textbf{Layers of}\\\textbf{Example}} \\
\midrule
Generation & Graph Topology &
\texttt{("topo-reshape-relu-shape", ["reshape", "relu"], ["reshape-out-shape", "relu-in-shape"], "\%reshape-out-shape EQUAL \%relu-in-shape")} & M \\
\cmidrule(l){2-4}
& Algebraic Semantics &
\texttt{("sem-matmul-k-match", ["matmul"], ["input1-width", "input2-height"], "\%input1-width EQUAL \%input2-height")} & M \\
\cmidrule(l){2-4}
& Operator Parameter Bound &
\texttt{("param-conv-stride-leq-kernel", ["conv"], ["stride-height", "kernel-height"], "\%stride-height LEQ \%kernel-height")} & M, C \\
\cmidrule(l){2-4}
& Data Representation Legality &
\texttt{("dtype-abs-supported-type-only", ["abs"], ["input-dtype"], "\%input-dtype IN \{FXP8, FXP16, FXP32\}")} & M, P \\
\cmidrule(l){2-4}
& Resource Compatibility &
\texttt{("res-abs-input-size", ["abs"], ["width", "height", "data-bits", "dram-size"], "\%width TIMES \%height TIMES \%data-bits LEQ \%dram-size")} & M, P \\
\midrule
Monitoring & Shape Consistency &
\texttt{"assert" OutputHeightBeforeFusionPass "==" OutputHeightAfterFusionPass} & M, C \\
\cmidrule(l){2-4}
& Hardware Alignment &
\texttt{"assert" PaddingFactor "\%" SystolicArrayM "==" "0"} & C, P \\
\cmidrule(l){2-4}
& Resource Capacity &
\texttt{"assert" TileSize "*" DataBits "<=" BufferCapacity} & M, C, P \\
\cmidrule(l){2-4}
& Transformation Result &
\texttt{"assert" LoopExtent "\%" TileWidth "==" "0"} & C \\
\cmidrule(l){2-4}
& Execution State &
\texttt{"assert" RuntimeState "!=" "ILLEGAL"} & C \\
\bottomrule
\end{tabular}
\endgroup
\end{table*}

\subsection{Testing DL Compilers}
We briefly describe the key components in testing DL compilers.
\begin{itemize}[leftmargin=*]
    \item {\bf Model Generation:} Testing DL compilers requires systematically generated DL models to exercise diverse compilation behaviors. Various model generators have been created that analyze syntactic and/or semantic validity constraints at the model or tensor level. They generate both valid models (which conform to type constraints) and invalid models (which intentionally violate constraints) to test compiler behaviors.
    For example, NNSmith~\cite{nnsmith} generates valid models by ensuring compatible tensor shapes, data types, and operator attributes, while it generates invalid models by violating these constraints.
    \item {\bf Test Oracle:} Prior work~\cite{neuri,hirgen,gencog} primarily relies on crash-based oracles and monitors coverage.
Another commonly used oracle~\cite{hirgen,nnsmith,neuri} takes model validity as the criterion: valid models are expected to compile, while invalid models should be rejected with an appropriate exception. Under this oracle, a compiler issue is reported if a valid model fails to compile or an invalid model is accepted without error.
\end{itemize}

Our extensive investigation of representative DL compilers~\cite{genesys,onnx-mlir,tvm,tensorrt,glow,tensorflow,pytorch} and model generation techniques~\cite{docter,nnsmith,neuri,hirgen,gencog,modelmeta,deeprel} reveals three key observations.

\MyPara{Observation 1:}
{\it Type-valid models do not always lead to successful compilation and hardware execution.}
For example, the GeneSys compiler~\cite{genesys} crashed while allocating FPGA memory for a valid model with tensor shapes \codefont{\small (1, 16, 10048934087, 22145136773)}. The failure occurred because the memory required for tensor allocation exceeded the available DRAM capacity. This suggests that compiler--platform interaction issues cannot be revealed by model validity alone, but require considering platform constraints across the compilation stack. Therefore, testing needs to incorporate such constraints when generating models.

\MyPara{Observation 2:}
{\it Cross-layer constraints can be mined but are largely overlooked in testing.}
Prior work has shown that DL operator constraints can be mined from model specifications and documentation~\cite{nnsmith,docter}.
We also observe that compilation- and hardware-related constraints are generally embedded in compiler documentation, compiler source code, and hardware specifications.
For example, ONNX-MLIR~\cite{onnx-mlir} implementation defines that the tiling factor for a model should be a multiple of the SIMD vector size, which can be obtained from hardware specifications.
Despite their importance, existing testing approaches focus mostly on operator-level constraints and overlook the compilation- and hardware-related constraints.

\begin{figure}[t]
  \centering
  \includegraphics{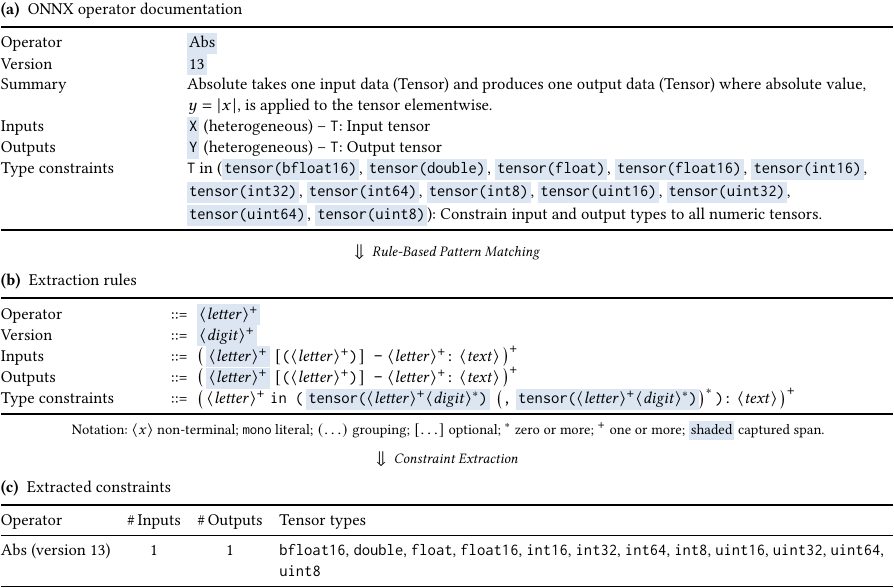}
  \caption{Constraint extraction from operator documentation.
(a)~Raw ONNX documentation for \texttt{Abs};
(b)~extraction rules, written in extended Backus--Naur form (EBNF), that match each documentation field;
(c)~structured constraints obtained by parsing.
Shaded spans mark the text captured in (a) and the rule symbols that capture it in (b);
the counts in (c) are the number of captured input and output tensor names.}
\label{fig:constraint-extraction}
\end{figure}

\MyPara{Observation 3:}
{\it Existing approaches observe a compilation only through its outcome and code coverage, which cannot differentiate compiler behaviors, while constraints naturally define finer-grained behavioral signals.}
In fact, a compilation behavior does not always manifest as an explicit signal such as a crash or a rejection. For example, in our experiments, two GeneSys tests went through the same padding pass and compiled without any error, while only one of them also padded the operands of a matrix multiplication to the 64-wide systolic array. In the end, both compilations succeeded, and neither the compilation outcome nor code coverage told them apart. Such behaviors differ in how intermediate values relate to the platform, such as the padding that the systolic array requires, but remain invisible to the compilation outcome and code coverage.

Based on these observations, we conclude that testing DL compilers requires analyzing constraints across the compilation stack, which goes beyond type validity at the model level. These observations form the foundation of \tool's cross-layer constraint extraction procedure, detailed in Section~\ref{sec:approach}.

\section{\large \tool}
\label{sec:approach}
\tool tests DL compilers that accept ONNX models.
Existing approaches build their tests mainly from model-level constraints, so the tests do not target the later stages where compiler--platform interaction issues arise, and each compilation is judged only as pass or fail.
\tool addresses this gap through two insights.
First, the conditions that decide whether a model gets through later stages, such as a tensor that must fit in device memory, are written down in operator documentation, hardware specifications, and compiler source code, so fixed extraction rules can turn them into cross-layer constraints.
Second, these constraints have two uses: generation constraints over quantities known before compilation determine how \tool builds a model that satisfies or breaks a chosen constraint, and monitoring constraints over values that the compiler computes determine what \tool records at the stage where the compiler computes these values.

Figure~\ref{fig:overview} outlines the workflow of \tool.
It first extracts cross-layer constraints from multiple sources and checks them before testing.
In each iteration, \tool then prioritizes a generation constraint and synthesizes a model that satisfies all applicable constraints or violates this constraint.
Finally, it compiles the model with guards that check the monitoring constraints, and the number of passes that the test completes updates the shallowness of the prioritized constraint for the next iteration.

\begin{figure*}[t]
  \centering
  \captionsetup[sub]{justification=centering}
  \definecolor{xcBlueBg}{RGB}{220,230,242}%
  \colorlet{xcBlue}{black}%
  \colorlet{xcPurple}{black}%
  \def\xcExt#1{{\setlength{\fboxsep}{0pt}\colorbox{xcBlueBg}{\rule[-2.3pt]{0pt}{8.6pt}\hspace{1pt}#1\hspace{1pt}}}}%
  \def\xcKw#1{\textcolor{xcPurple}{#1}}%
  \newif\ifxcCont
  \def\xcWrap{\global\xcConttrue}%
  \def\xcNum#1{\ifxcCont\global\xcContfalse\addtocounter{lstnumber}{-1}\else#1\fi}%
  \lstdefinestyle{xcRuleTpl}{
    language={}, backgroundcolor={},
    basicstyle=\ttfamily\fontsize{7}{9.5}\selectfont,
    keywordstyle={}, stringstyle={}, commentstyle={}, identifierstyle={}, emphstyle={},
    keywords={}, otherkeywords={}, emph={}, linewidth=\linewidth, escapechar={},
    numbers=left, numberstyle=\fontsize{5}{6}\selectfont\color{gray}\xcNum,
    numbersep=3pt, stepnumber=1, firstnumber=1, numberblanklines=true,
    breaklines=false, columns=fullflexible, keepspaces=true,
    showstringspaces=false, showlines=true,
    frame=tb, framesep=3pt, framerule=0.4pt, rulecolor=\color{black},
    xleftmargin=8pt, framexleftmargin=8pt, xrightmargin=0pt, framexrightmargin=0pt,
    aboveskip=0pt, belowskip=0pt, escapeinside={(*}{*)},
    literate={->}{{\textcolor{xcBlue}{->}}}2,
  }%
  \begin{subfigure}[t]{0.35\textwidth}
\begin{lstlisting}[style=xcRuleTpl]
ArrayHeight -> (*\xcExt{ExtArrayHeightKey}\xcWrap*)
    "=" Number
ArrayWidth -> (*\xcExt{ExtArrayWidthKey}\xcWrap*)
    "=" Number
DataSize -> (*\xcExt{ExtDataSizeKey}\xcWrap*)
    "=" Number
Bandwidth -> (*\xcExt{ExtBandwidthKey}\xcWrap*)
    "=" Number

(*\xcExt{ExtArrayHeightKey}*) -> (*\xcKw{"ARRAY\_M"}*) | ...
(*\xcExt{ExtArrayWidthKey}*) -> (*\xcKw{"ARRAY\_N"}*) | ...
(*\xcExt{ExtDataSizeKey}*) -> (*\xcKw{"DATA\_WIDTH"}*) | ...
(*\xcExt{ExtBandwidthKey}*) -> (*\xcKw{"IBUF\_CHANNEL\_BW"}\xcWrap*)
    | ...(*\xcWrap*)

\end{lstlisting}
    \caption{\bf Extraction Rules for HW Specs.}
    \label{fig:hw-spec-rule-template}
  \end{subfigure}%
  \hfill
  \begin{subfigure}[t]{0.32\textwidth}
\begin{lstlisting}[style=xcRuleTpl]
PaddingFactor -> VarName(*\xcWrap*)
    "=" PFCall
PFCall -> (*\xcExt{ExtPadFactorOpName}\xcWrap*)
    "(" VarName "," VarName ")"


NewTensorShape -> VarName(*\xcWrap*)
    "=" PadCall
PadCall -> (*\xcExt{ExtPadOpName}\xcWrap*)
    "(" VarName "," VarName ")"


(*\xcExt{ExtPadFactorOpName}*) -> (*\xcKw{"lcm"}*) | ...
(*\xcExt{ExtPadOpName}*) -> (*\xcKw{"pad\_fn"}*) | ...(*\xcWrap*)

\end{lstlisting}
    \caption{\bf Extraction Rules for Compilers.}
    \label{fig:impl-rule-template}
  \end{subfigure}%
  \hfill
  \begin{subfigure}[t]{0.31\textwidth}
\begin{lstlisting}[style=xcRuleTpl]
ArraySize -> ArrayHeight(*\xcWrap*)
    "*" ArrayWidth

PaddingFactorConstraint ->
    "assert" PaddingFactor "%"(*\xcWrap*)
        Bandwidth "==" "0"
    "assert" PaddingFactor "%"(*\xcWrap*)
        ArraySize "==" "0"

NewTensorWidthConstraint ->
    "assert" NewTensorWidth "%"(*\xcWrap*)
        PaddingFactor "==" "0"
NewTensorHeightConstraint ->
    "assert" NewTensorHeight "%"(*\xcWrap*)
        PaddingFactor "==" "0"
\end{lstlisting}
    \caption{\bf Conditions as Guards for Padding.}
    \label{fig:compilation-constr-template}
  \end{subfigure}
  \caption{Extraction rules for hardware specifications and compiler code, and the assertion templates that turn the extracted padding values into guards.}
  \label{fig:compilation-constraint-ex}
\end{figure*}

\subsection{Deterministic Cross-Layer Constraint Extraction and Checking}
\label{sec:constraint-inference}

\tool extracts constraints with fixed extraction rules, which match structured documentation, hardware specifications, and ASTs of compiler code.
Such extraction is deterministic, as each rule matches fixed patterns without any random choice or learned model (such as large language models, LLMs). Each constraint therefore follows directly from the source text or code that states it, so, unlike constraints that an LLM infers, it needs no separate check of whether it is correct.

\MyPara{Cross-Layer Constraints and Their Uses.}
The extracted constraints relate quantities from three layers of the compilation stack.
Quantities of the model include graph connectivity, operator semantics, tensor shapes, data types, and attributes; quantities of the compiler include intermediate representations and the results of compilation passes; and quantities of the platform include supported representations, memory capacities, bandwidth, and array dimensions.
A constraint that relates quantities from two or three layers is a cross-layer constraint.
Besides the layers it relates, each constraint has one of two uses: a {\it generation constraint} involves only quantities known before compilation and guides model synthesis, whereas a {\it monitoring constraint} involves values that the compiler computes and is checked by a guard during compilation.

\MyPara{Constraint Representation.}
\tool represents generation and monitoring constraints differently.
A generation constraint has the form
\begin{center}
\codefont{("identifier", ["op1", ...], ["arg1", ...], "relation")}.
\end{center}
Its identifier names the rule, the operator list determines where it applies, and the argument list names the quantities in the relation.
For example, a \codefont{Reshape} followed by a \codefont{Relu} must pass on a tensor of the same shape, so the constraint \codefont{topo-reshape-relu-shape} requires the shape of the output tensor of \codefont{Reshape} to equal the shape of the input tensor of \codefont{Relu}.
A monitoring constraint is represented as \codefont{assert} followed by a predicate over the values available at a compilation checkpoint.

We group the constraints into ten categories by what they describe, listed in Table~\ref{tab:constr-category}.
Each row gives an example of its category in the representation of \tool, together with the layers that the example relates.

Five categories of generation constraints guide model synthesis: graph topology, algebraic semantics, operator parameter bounds, data representation legality, and resource compatibility.
For example, the input tensor of \codefont{Abs} must fit in the DRAM of the platform, so \codefont{res-abs-input-size} requires its width times its height times its bit width to be at most the DRAM size, and it relates the model (M) to the platform (P).
In addition, five categories of monitoring constraints describe intermediate compilation states: shape consistency, hardware alignment, resource capacity, transformation results, and execution states.
For example, a tile that the compiler keeps in an on-chip buffer must fit in that buffer, so the resource-capacity constraint requires the tile size that the compiler selects for a dimension of a model tensor, times the bit width of the tensor's data type, to be at most the buffer capacity, and it relates all three layers (M, C, P).

\begin{figure}[t]
  \centering
  \includegraphics{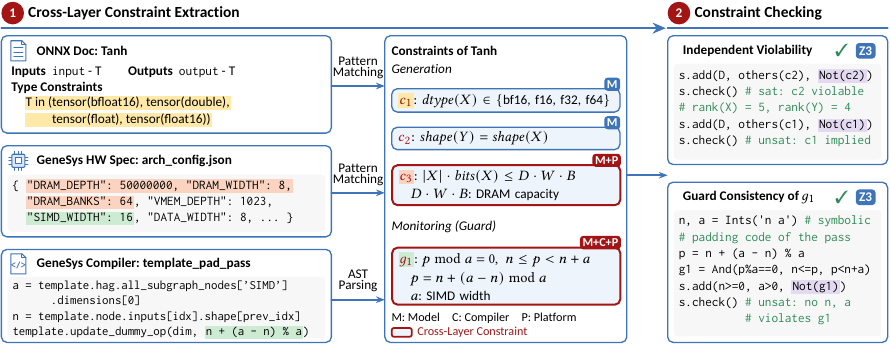}
  \caption{A working example of cross-layer constraint extraction and checking for \codefont{Tanh} on GeneSys.
(1) \tool extracts the generation constraints $c_1$ and $c_2$ from the ONNX documentation, the generation constraint $c_3$ from the documentation and the hardware specification, and the padding guard $g_1$ from the compiler code.
(2) Z3 shows that $c_2$ can be violated alone while $c_1$ follows from the other constraints, and that no symbolic $n\geq 0$ and $a>0$ violates $g_1$, so $g_1$ holds for every dimension size $n$ and SIMD width $a$; the highlighted \codefont{Not} marks the negated constraint in each query.}
  \label{fig:example-extraction-checking}
\end{figure}

\MyPara{Pattern Matching in Documentation.}
Operator documentation states the constraints of each operator explicitly.
It also follows a consistent structure, so \tool encodes one set of extraction rules for each documentation format.
Take ONNX documentation as an example.
Figure~\ref{fig:constraint-extraction}(a) shows the page of \codefont{Abs}.
It begins with the operator name and its version, followed by a summary and the sections \codefont{Inputs}, \codefont{Outputs}, and \codefont{Type constraints}.
Every operator page follows the same structure.
Figure~\ref{fig:constraint-extraction}(b) gives one extraction rule for each of these fields.
The rules are written in extended Backus--Naur form (EBNF)~\cite{iso14977}, a standard notation for grammars.
Each rule matches one field of the documentation and extracts its values, such as the names of the input tensors or the types that a type constraint lists.
For example, the \codefont{Inputs} rule matches each entry that starts with a tensor name, such as \codefont{X}, followed by its type parameter \codefont{T}.
Counting its matches shows that \codefont{Abs} has one input tensor.
The \codefont{Outputs} rule finds the output tensor \codefont{Y} in the same way.
Similarly, the \codefont{Type constraints} rule captures every \codefont{tensor(...)} listed after \codefont{T in}, which gives 12 types from \codefont{tensor(bfloat16)} to \codefont{tensor(uint8)}.
Figure~\ref{fig:constraint-extraction}(c) lists these results.
From them, \tool instantiates constraints such as ``\codefont{Abs} has exactly one input tensor'' and ``the type of \codefont{X} is one of the 12 types''.

\MyPara{Pattern Matching in Hardware Specifications.}
The hardware specifications of different platforms describe common parameters, such as array dimensions, data width, and bandwidth.
However, different platforms name these parameters differently.
For example, GeneSys names the height of its systolic array \codefont{ARRAY\_M}, whereas Timeloop-style specifications~\cite{timeloop} name it \codefont{meshX}.
To handle such formats with the same rules, we write the extraction rules in Figure~\ref{fig:hw-spec-rule-template} over abstract keywords.
These keywords are the nonterminals that start with \codefont{Ext}, and we instantiate them for each platform.
For example, Line~1 retrieves the array height with \codefont{ArrayHeight -> ExtArrayHeightKey "=" Number}.
Line~6 instantiates \codefont{ExtArrayHeightKey} with \codefont{ARRAY\_M} for GeneSys, and a Timeloop-style specification only needs \codefont{meshX} as another instance.
The matched values become concrete platform parameters.
Together with the quantities of the model, they instantiate capacity and alignment constraints, such as the DRAM-capacity constraint of \codefont{Abs}.

\MyPara{AST Parsing in Compiler Implementation.}
Unlike documentation, compiler code states the conditions of a transformation implicitly, through the values that the transformation computes.
\tool therefore matches extraction rules against the ASTs of compiler code.
We currently target four commonly used transformations, namely tiling, padding, loop unrolling, and operator fusion, as introduced in Table~\ref{tab:hardware}.
The rules locate these transformations in the compiler code and identify the variables that each transformation computes and the locations where its conditions can be checked.
Take padding as an example.
A padding transformation first computes a padding factor from the platform.
It then pads a tensor dimension to a multiple of this factor.
The extraction rules in Figure~\ref{fig:impl-rule-template} locate these two computations through the keywords \codefont{ExtPadFactorOpName} and \codefont{ExtPadOpName} (Lines~9 and 10).
For GeneSys, these keywords are instantiated with \codefont{lcm} and \codefont{pad\_fn}, so the rules find the statements that compute the padding factor and the padded shape.
The assertion templates in Figure~\ref{fig:compilation-constr-template} then write the padding conditions over the matched variables.
For example, \codefont{assert NewTensorWidth \% PaddingFactor == 0} requires the padded width to be a multiple of the padding factor.

\begin{table}[t]
\begin{algorithm}[H]\small
\caption{Constraint Prioritization}
\label{alg:prioritization}
\KwIn{Constraint profiles $P=\{(c_i,s_i,u_i)\}$ with shallowness $s_i$ and usage count $u_i$; generation constraints $C$; platform configuration $h$; testing budget $T$}
\KwOut{Updated constraint profiles $P$}
\While{the testing budget $T$ remains}{
  $S \gets$ constraints with the lowest shallowness in $P$\;
  $c_k \gets$ a least frequently used constraint in $S$\;
  $u_k \gets u_k+1$\tcp*{one more use of $c_k$}
  $m \gets$ draw a valid or targeted-violation mode\tcp*{satisfy or violate $c_k$}
  $(M,h) \gets$ synthesize a model using $(C,c_k,h,m)$\;
  \uIf{synthesis succeeds}{
    Compile $M$ for $h$ with guards in place\;
    $p \gets$ number of passes that the test completes\;
    $s_k \gets s_k + 1/\max(1,p)$\;
  }
  \Else{
    $s_k \gets s_k+1$\;
  }
}
\Return $P$\;
\end{algorithm}

\vspace{1.5em}
\centering
\small
\caption{Working example of constraint prioritization. Shallowness values and usage counts are those before each iteration. Min.\ shallowness prioritizes the constraint with the lowest shallowness, LFU prioritizes the least frequently used constraint among those with the lowest shallowness, and Random picks one of the remaining constraints at random. Over the nine iterations, \tool prioritizes \codefont{C2}, whose tests complete the most passes, four times, against three times for \codefont{C3} and twice for \codefont{C1}, so its tests focus on the deeper logic of the compiler.}
\label{tab:prioritization-ex}
\begingroup
\setlength{\tabcolsep}{3pt}
\renewcommand{\arraystretch}{1.15}
\begin{tabular*}{\linewidth}{@{\extracolsep{\fill}}ccccccl@{}}
\toprule
\multicolumn{7}{c}{Passes that each test completes before failure:\quad \codefont{C1}: 1\qquad \codefont{C2}: 4\qquad \codefont{C3}: 3} \\
\midrule
\textbf{Iter.} & \makecell{\textbf{Shallowness}\\$(s_1,s_2,s_3)$} & \makecell{\textbf{Usage Counts}\\$(u_1,u_2,u_3)$} & \textbf{Prioritized} & \textbf{Criterion} & \makecell{\textbf{Passes}\\$p$} & \makecell[l]{\textbf{Shallowness}\\\textbf{Update}} \\
\midrule
1 & $(1,1,1)$ & $(0,0,0)$ & \codefont{C1} & Random & 1 & $1+1/1=2$ \\
2 & $(2,1,1)$ & $(1,0,0)$ & \codefont{C2} & Random & 4 & $1+1/4=1.25$ \\
3 & $(2,1.25,1)$ & $(1,1,0)$ & \codefont{C3} & Min. shallowness & 3 & $1+1/3=4/3$ \\
4 & $(2,1.25,4/3)$ & $(1,1,1)$ & \codefont{C2} & Min. shallowness & 4 & $1.25+1/4=1.5$ \\
5 & $(2,1.5,4/3)$ & $(1,2,1)$ & \codefont{C3} & Min. shallowness & 3 & $4/3+1/3=5/3$ \\
6 & $(2,1.5,5/3)$ & $(1,2,2)$ & \codefont{C2} & Min. shallowness & 4 & $1.5+1/4=1.75$ \\
7 & $(2,1.75,5/3)$ & $(1,3,2)$ & \codefont{C3} & Min. shallowness & 3 & $5/3+1/3=2$ \\
8 & $(2,1.75,2)$ & $(1,3,3)$ & \codefont{C2} & Min. shallowness & 4 & $1.75+1/4=2$ \\
9 & $(2,2,2)$ & $(1,4,3)$ & \codefont{C1} & LFU & -- & -- \\
\bottomrule
\end{tabular*}
\endgroup
\end{table}

\MyPara{Constraint Refinement.}
The extracted constraints describe an operator in general, but a compiler may support only part of it.
\tool therefore cross-references the extracted constraints with compiler support and the platform configuration to refine where each constraint applies.
For example, GeneSys selects a codelet for \codefont{Add} only for the ranks of input tensors listed in its registry, so the rank constraints of \codefont{Add} on GeneSys are narrower than those of the ONNX documentation.

\MyPara{Constraint Checking.}
Although the extraction is deterministic, a constraint that is correct on its own can still be unusable for testing.
First, one constraint of an operator can follow from its other constraints, so no model violates it alone.
Second, a guard is written from a general template rather than from the code of a specific pass, so its condition may be stricter than what the pass computes.
Such a guard would fail even when the compiler is correct.
\tool therefore checks each generation constraint for independent violability and each guard for consistency with its transformation.
Both checks run once before testing and need no compiler.

For independent violability, \tool encodes the generation constraints of each operator as Z3~\cite{z3} formulas over symbolic ranks, dimensions, data types, and attribute values.
For each constraint, Z3 searches for an assignment that makes this constraint false and keeps the other constraints true.
If no such assignment exists, the constraint is implied by the others.
\tool keeps an implied constraint for valid models but never targets it for violation.

For guard consistency, we check that a correct pass never fails its guard: for every input that the pass receives, the value that it computes satisfies the guard.
Z3 proves an arithmetic guard when it can decide the guard over unbounded integers, such as the padding guard in Figure~\ref{fig:example-extraction-checking}.
When the computation divides by a variable, Z3 does not decide the guard within its time limit, and we prove the guard in Lean~\cite{lean4} instead.
For example, a padding pass of GeneSys rounds a dimension size $n$ up to $\lceil n/a\rceil\cdot a$, and Lean proves that the result is a multiple of $a$, at least $n$, and less than $n+a$ for all natural numbers $n$ and $a>0$.
The remaining guards need no arithmetic proof.
Each of them either states a constraint of the platform directly, such as a guard that checks whether a tile that the compiler selects fits in its on-chip buffer, or restates the definition of a transformation, such as a guard that checks whether an unrolled loop holds one copy of its body per iteration.
We manually review these guards.

\begin{algorithm}[t]\small
\caption{Model Synthesis}
\label{alg:model-synthesis}
\KwIn{Generation constraints $C$; prioritized constraint $c_k$; platform configuration $h$; test mode $m$}
\KwOut{An ONNX model $M$ with configuration $h$, or synthesis failure}
\For{each synthesis attempt}{
  $G \gets$ construct a graph around the operator of $c_k$\;
  $C_G \gets$ apply the constraints of each node of $G$ to its tensors and $h$\tcp*{$c_k^G$: $c_k$ on the placed node}
  $\Phi \gets \bigwedge_{c\in C_G\setminus\{c_k^G\}} c$\tcp*{all other constraints hold}
  $\Phi \gets \Phi \land \mbox{target\_relation}(c_k^G,m)$\tcp*{$c_k^G$ if valid, $\neg c_k^G$ otherwise}
  $\theta \gets$ solve $\Phi$\;
  \If{an assignment $\theta$ is found}{
    $M \gets$ save $(G,\theta)$ as an ONNX model and read it back\;
    \If{the saved model $M$ satisfies $\Phi$}{
      \Return $(M,h)$\;
    }
  }
}
\Return synthesis failure\;
\end{algorithm}

\MyPara{Example.}
Figure~\ref{fig:example-extraction-checking} puts these steps together for one operator of GeneSys, \codefont{Tanh}, to demonstrate how \tool extracts its constraints from all three sources and checks them.
From the ONNX documentation, \tool extracts two generation constraints.
The type section of \codefont{Tanh} lists \codefont{bfloat16}, \codefont{double}, \codefont{float}, and \codefont{float16}.
The type constraint ($c_1$) therefore requires the input tensor $X$ to have one of these four types.
The documentation also states that \codefont{Tanh} is applied element-wise.
The shape constraint ($c_2$) therefore requires the output tensor $Y$ to have the shape of $X$.
From the hardware specification \codefont{arch\_config.json}, \tool reads \codefont{DRAM\_DEPTH}~$D=50{,}000{,}000$, \codefont{DRAM\_WIDTH}~$W=8$, and \codefont{DRAM\_BANKS}~$B=64$.
Their product, $25.6\times 10^{9}$ bits, is the DRAM capacity.
The DRAM-capacity constraint ($c_3$) requires the number of elements of $X$ times its bit width to stay within this capacity, so it relates the model to the platform.
From the GeneSys compiler, AST parsing finds the statement of \codefont{template\_pad\_pass} that pads a dimension of an input tensor from size $n$ to size $p=n+(a-n)\bmod a$.
Here, $a$ is the width of the SIMD unit.
The padding guard ($g_1$) requires $p\bmod a=0$ and $n\leq p<n+a$, so it relates a dimension of the model, a value that the compiler computes, and a parameter of the platform.
The variables of $c_1$--$c_3$ are known before compilation, so these three constraints are generation constraints.
In contrast, $p$ exists only inside the compiler, so $g_1$ checks a monitoring constraint.
Z3 then shows that $c_2$ can be violated alone: an output tensor whose rank differs from that of $X$ breaks $c_2$, while the other constraints still hold.
In contrast, $c_1$ is implied, since the output type must lie in the same four types and equal the input type.
Z3 also finds no symbolic $n\geq 0$ and $a>0$ that violates $g_1$, so $g_1$ holds for every dimension size and SIMD width.

\subsection{Constraint-Guided Model Synthesis}
\label{sec:model-synthesis}

Testing DL compilers needs both valid and invalid models~\cite{nnsmith,neuri,docter}.
A valid model satisfies all constraints, but an invalid model must break some constraint, and the choice of this constraint matters.
Breaking an arbitrary constraint mostly leads to a rejection at the model validity check.
For example, a model with mismatched tensor shapes is rejected right away.
In contrast, a model with valid shapes but a tensor that exceeds the DRAM capacity passes the model validity check and reaches the later stages that allocate memory.
Valid models face a similar choice, since the constraint that a model is built around decides which operators and transformations the model exercises.

We address this challenge with an idea inspired by feedback-guided fuzzing, which keeps the inputs that reach new program branches~\cite{afl} and favors branches that are rarely hit~\cite{fairFuzz}.
\tool treats each generation constraint as a target and uses compilation feedback in two ways.
First, it records how many compilation passes the tests of each constraint complete and prefers constraints whose tests reach later stages.
Second, among constraints whose tests reach similar stages, it prefers the least frequently used ones to spread testing across constraints.
In each iteration, \tool therefore first prioritizes a generation constraint and then generates a model for this constraint, as Algorithm~\ref{alg:prioritization} shows.

\begin{figure}[t]
  \centering
  \includegraphics{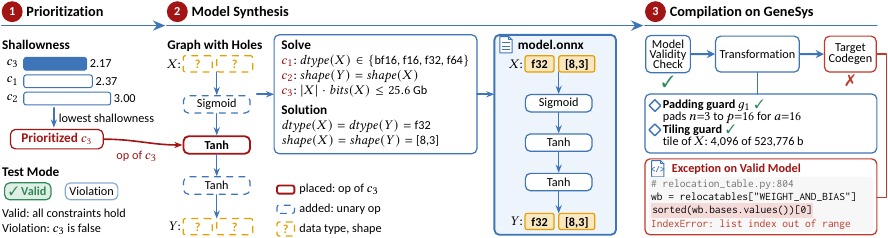}
  \caption{A working example of constraint-guided model synthesis and compilation, following Figure~\ref{fig:example-extraction-checking}.
(1) \tool prioritizes $c_3$, the constraint with the lowest shallowness, and requests a valid model.
(2) \tool places a \codefont{Tanh} node, the operator of $c_3$, and adds unary operators that keep the shape and data type.
It then solves the constraints of all three nodes for the data types and shapes of $X$ and $Y$.
(3) GeneSys raises an \codefont{IndexError} in code generation, because it indexes the empty weight region of this valid model.
Before this exception, the model passes the earlier stages, where the padding guard $g_1$ and a tiling guard hold.}
  \label{fig:example-synthesis-compilation}
\end{figure}

\MyPara{Constraint Prioritization.}
Algorithm~\ref{alg:prioritization} is the main testing loop of \tool.
It keeps a shallowness $s_i$ and a usage count $u_i$ for each generation constraint $c_i$, and every shallowness starts at 1.
The shallowness of a constraint grows faster when the tests of this constraint stop at earlier compilation stages.
In each iteration, \tool prioritizes the constraint with the lowest shallowness (Lines~2--3).
If several constraints share the lowest shallowness, it prioritizes the one that has been used least often, \ie\ the least frequently used (LFU) constraint, and if they also share the usage count, it picks one of them at random.
\tool then decides whether to generate a valid or a targeted-violation model for this constraint (Line~5), synthesizes the model (Line~6), and compiles the model with the guards in place (Line~8).

The shallowness update turns compilation feedback into priority: the further the tests of a constraint get, the more slowly its shallowness grows, so the constraint is prioritized again sooner.
Inspired by spectrum-based fault localization~\cite{tarantula,ochiai,hsbfl}, which scores each program element by the outcomes of the tests that execute it, \tool scores each constraint by how far its tests get in the compiler.
Specifically, it counts the number $p$ of passes that the test completes before its compilation finishes or fails, and updates the shallowness of the prioritized constraint as
\[
  s_k \gets s_k + \frac{1}{\max(1,p)}.
\]
For example, a test that completes 20 passes adds only 0.05 to the shallowness, so its constraint stays among the first choices.
In contrast, a test that stops after the first pass adds 1, and so does a test that fails before any pass, since $\max(1,p)$ is then 1.
The same penalty applies when synthesis fails (Line~12).
A constraint for which \tool cannot build a model reaches no compilation stage at all, so it should give way to other constraints at least as quickly as a constraint whose tests stop at once.
\tool repeats the loop until the testing budget is exhausted.

Table~\ref{tab:prioritization-ex} walks through Algorithm~\ref{alg:prioritization} in nine iterations over three constraints, \codefont{C1}, \codefont{C2}, and \codefont{C3}, whose tests complete one, four, and three passes before failing.
All shallowness values start at 1 and no constraint has been used, so the first two iterations pick at random and prioritize \codefont{C1} and then \codefont{C2}.
\codefont{C1} completes only one pass, so its shallowness rises to 2, and it is not prioritized again until iteration 9.
Each test of \codefont{C2} and \codefont{C3} adds only $1/4$ or $1/3$, so the two constraints take turns as the constraint with the lowest shallowness in iterations 3 to 8.
Once all three shallowness values equal 2, LFU prioritizes \codefont{C1} in iteration 9, since it has been used once, against four times for \codefont{C2} and three times for \codefont{C3}.
As a result, \codefont{C2} and \codefont{C3}, whose tests get further into the compiler, take up iterations 2 to 8, while LFU still returns to \codefont{C1} so that no constraint is left out of testing.

\MyPara{Model Generation.}
\tool supports two test modes for a prioritized constraint $c_k$.
In the valid mode, the model satisfies every applicable generation constraint.
In the targeted-violation mode, the model makes $c_k$ false and keeps the other constraints true.

Given the prioritized constraint $c_k$ and a test mode, \tool generates a test $t=(G,\theta,h)$, which consists of the computation graph $G$ of a model, an assignment $\theta$ of its tensor shapes, data types, and operator attributes, and a platform configuration $h$, such as the memory sizes and array dimensions of an accelerator.
Algorithm~\ref{alg:model-synthesis} shows the procedure.
\tool first places a node of the operator that $c_k$ constrains and extends the graph up to five nodes with unary operators that keep the shape and data type, such as \codefont{Relu}, \codefont{Sigmoid}, and \codefont{Identity} (Line~2).
The extracted constraints of an operator refer to its tensors in general, such as the input tensor $X$ of \codefont{Tanh}, rather than to a node of $G$.
\tool therefore applies the constraints of each node of $G$ to the tensors of this node, which gives the set $C_G$ (Line~3).
A graph can contain several nodes of the same operator, such as \codefont{Sigmoid}$\rightarrow$\codefont{Tanh}$\rightarrow$\codefont{Tanh}, so $C_G$ can contain several copies of $c_k$.
The copy on the node that \tool places for $c_k$ is denoted $c_k^G$, and a targeted-violation model makes only this copy false.
Constraints that involve the platform also read fields of $h$, such as the DRAM capacity.
Next, \tool solves for the shapes, data types, and attributes (Lines~4--7).
For a valid model, it solves for parameters that satisfy all constraints.
For a targeted-violation model, it changes this valid model around $c_k$, such as one attribute value or one dimension, until $c_k$ is false and the other constraints hold.
Finally, \tool keeps the model only if the saved ONNX file still satisfies the requested mode (Lines~8--9).

\MyPara{Example.}
Figure~\ref{fig:example-synthesis-compilation} continues with the constraints of \codefont{Tanh} on GeneSys to demonstrate how \tool prioritizes one of them, synthesizes a model for it, and compiles this model with guards.
Constraint $c_3$ has shallowness 2.17, lower than 2.37 for $c_1$ and 3.00 for $c_2$, so \tool prioritizes $c_3$ and requests a valid model.
\tool places a \codefont{Tanh} node and attaches \codefont{Sigmoid} before it and a second \codefont{Tanh} after it.
It then solves the constraints of all three nodes for the data types and shapes of the input tensor $X$ and output tensor $Y$ of the model, which gives the type \codefont{float32} and the shape $[8,3]$.
For example, $X$ needs $8\cdot 3\cdot 32=768$ bits, within the $25.6\times 10^{9}$ bits of DRAM, so $c_3$ holds.
The model is thus ready to be compiled with guards on GeneSys.

\subsection{Guard-Based Compilation Monitoring}
\label{sec:behavior-diff}

Existing DL compiler testing approaches distinguish compilation behaviors by coarse-grained signals, such as crashes and compilation results.
These signals often merge different behaviors into the same outcome.
For example, a compilation may fail because a tensor does not fit in memory or because the compiler computes a wrong tile size.
Both failures can show up as the same crash when backend lowering allocates buffers.
Coverage-based metrics, such as branch coverage and pass coverage, have a similar limitation.
They treat two compilations as the same as long as both reach the same code and passes.
For example, two models can execute the same padding code, while one of them pads a dimension and the other needs no padding because its dimension is already aligned.
Both compilations follow the same control flow, so their branch and pass coverage are identical.
\tool therefore inserts guards where the compiler computes such values and records what each guard observes.

\MyPara{Guard Insertion.}
\tool inserts the extracted guards into the compiler source as assertions.
Since the extraction rules target the four commonly used transformations in Table~\ref{tab:hardware}, namely tiling, padding, loop unrolling, and operator fusion, the guards sit in the code of these transformations in each compiler.
Each assertion is placed right after the transformation computes the values that the guard reads.
A guard only reads these values and does not change the intermediate representation or any decision of the compiler.
The guards relate these values to quantities of the platform or of the transformation itself.
For example, a resource-capacity guard relates a tile to the capacity of its buffer, a hardware-alignment guard relates a padded dimension size to the alignment of the platform, and a transformation-result guard relates the tiles of a loop to its number of iterations.

Take the padding pass of GeneSys in Figure~\ref{fig:assertion-ex} as an example.
The compiler computes the padding factor \codefont{constr} as the least common multiple of \codefont{bandwidth} and \codefont{sys\_dims}.
\codefont{pad\_fn} then rounds the dimension \codefont{dummy\_inp\_dim} of the input tensor up to \codefont{new\_inp\_dim}, a multiple of this factor.
\tool inserts four guards right after this computation, shown as the lines marked \codefont{+}.
The first guard requires the padding factor to be positive.
The other three come from the padding guard $g_1$.
They require the padded size to be a multiple of the padding factor, to be no smaller than the original size, and to exceed the original size by less than one padding factor.

\begin{figure}[t]
  \centering
  \definecolor{xcGuardRed}{HTML}{9B1C1C}%
  \lstdefinestyle{xcGuardDiff}{
    language={}, keywords={}, otherkeywords={}, emph={}, literate={},
    basicstyle=\ttfamily\small, backgroundcolor={},
    keywordstyle={}, stringstyle={}, commentstyle={}, identifierstyle={}, emphstyle={},
    numbers=none, breaklines=false, escapechar={}, escapeinside={},
    columns=fullflexible, keepspaces=true, showspaces=false, showtabs=false,
    showstringspaces=false,
    frame=tb, framesep=3pt, framerule=0.4pt, rulecolor=\color{black},
    xleftmargin=0pt, xrightmargin=0pt, framexleftmargin=0pt, framexrightmargin=0pt,
    linewidth=\linewidth, aboveskip=0pt, belowskip=0pt,
    moredelim=[l][\color{xcGuardRed}]{+\ },
  }%
  \begin{minipage}{0.62\textwidth}
\begin{lstlisting}[style=xcGuardDiff]
  constr = np.lcm(bandwidth, sys_dims)
  pad_fn = lambda shape, v: -1*(shape // (-1*v))*v
  dummy_inp_dim = template.node.inputs[0].shape[1]
  new_inp_dim = pad_fn(dummy_inp_dim, constr)
+ guard(constr > 0)
+ guard(new_inp_dim % constr == 0)
+ guard(new_inp_dim >= dummy_inp_dim)
+ guard(new_inp_dim - dummy_inp_dim < constr)
\end{lstlisting}
  \end{minipage}
  \caption{Guard insertion in the padding pass of GeneSys. Lines marked with \texttt{+} are inserted by \tool.}
  \label{fig:assertion-ex}
\end{figure}

\MyPara{Behavioral Observations.}
Each time a compilation reaches a guard, \tool records that the guard is exercised, together with the operator that it checks, the values that it reads, and whether its predicate holds.
The guards that a compilation exercises thus show what the compiler computes along the way.
For example, two models can reach the same padding pass and exercise the same padding guard, but the guard records that the dimension of one model is padded from 3 to 16 and that the dimension of the other is already aligned.
These observations complement the pass counts that guide constraint prioritization.

\MyPara{Example.}
Figure~\ref{fig:example-synthesis-compilation} also shows how GeneSys compiles the model that \tool synthesizes for $c_3$, to demonstrate what the guards record.
After the model validity check, GeneSys pads the last dimension of $X$ from 3 to 16 to match the SIMD width of 16, so the padding guard $g_1$ holds.
The tiling guard then records that the tile of $X$ needs 4,096 bits, within the 523,776 bits of its buffer.
In code generation, GeneSys looks up the empty weight region of the model and raises an \codefont{IndexError}.
Without guards, this test looks the same as any other test that fails in code generation.
The guards provide an additional behavioral signal: they record that this compilation padded $X$ and kept its tile within the buffer before the exception.
This signal distinguishes the test from other tests that fail at the same point after different padding or tiling.

\section{Evaluation Results}
\label{sec:evaluation}
\pgfplotsset{compat=1.17}
\definecolor{catSC}{HTML}{1F3A93}
\definecolor{catSI}{HTML}{2E8B57}
\definecolor{catIV}{HTML}{E8743B}
\definecolor{catDT}{HTML}{F2B134}
\definecolor{catPI}{HTML}{B7D8F2}
\definecolor{catIL}{HTML}{8E7CC3}
\definecolor{stageCG}{HTML}{1F4E8C}
\definecolor{stageBL}{HTML}{4F8FCB}
\definecolor{stageGT}{HTML}{A6C8E8}
\definecolor{stageMV}{HTML}{EEF1F5}
\definecolor{heatRQ1}{HTML}{3E7CB1}
\definecolor{heatRQ4}{HTML}{00796B}
\definecolor{apXC}{HTML}{0072B2}
\definecolor{apXU}{HTML}{56B4E9}
\definecolor{apNR}{HTML}{E69F00}
\definecolor{apDR}{HTML}{009E73}
\definecolor{apGC}{HTML}{CC79A7}
\definecolor{apHG}{HTML}{D55E00}
\pgfplotsset{
  eval axis/.style={
    font=\footnotesize,
    tick label style={font=\footnotesize},
    label style={font=\footnotesize},
    legend style={font=\footnotesize, draw=none, fill=none},
    axis line style={black!70},
    tick style={black!70},
    grid style={black!12},
  },
}

\providecommand{\rqsummary}[2]{%
\begin{tcolorbox}[colback=black!4, colframe=black!65, boxrule=0.5pt, arc=2pt,
  left=5pt, right=5pt, top=3pt, bottom=3pt,
  fonttitle=\bfseries\small, title={Summary of RQ#1}]
#2
\end{tcolorbox}}

We seek to answer the following research questions.
\begin{description}[leftmargin=2.8em,style=nextline]
    \item[RQ1] What cross-layer constraints does \tool extract, and how reliable are they?
    \item[RQ2] What symptoms does \tool detect in DL compilers, and what are their root causes?
    \item[RQ3] How effectively does \tool reach later compilation stages?
    \item[RQ4] How effectively do the guards of \tool observe intermediate compilation states?
\end{description}

\definecolor{RQxcheck}{HTML}{3C5488}
\definecolor{RQneuri}{HTML}{B09C85}
\definecolor{RQgencog}{HTML}{00A087}
\definecolor{RQhirgen}{HTML}{4DBBD5}
\definecolor{RQdeeprel}{HTML}{F39B7F}
\pgfplotstableread{figures/data/rq1-constraint-layers.dat}\RQtab
\pgfplotstablegetrowsof{\RQtab}\pgfmathtruncatemacro\RQn{\pgfplotsretval-1}%
\pgfplotsforeachungrouped \i in {0,...,4}{%
  \expandafter\xdef\csname RQmax\i\endcsname{0}%
  \expandafter\xdef\csname RQkind\i\endcsname{0}}%
\pgfplotsforeachungrouped \i in {0,...,\RQn}{%
  \pgfplotstablegetelem{\i}{x}\of\RQtab \pgfmathtruncatemacro\RQx{\pgfplotsretval}%
  \pgfplotstablegetelem{\i}{y}\of\RQtab \pgfmathtruncatemacro\RQy{\pgfplotsretval}%
  \pgfplotstablegetelem{\i}{count}\of\RQtab \let\RQc\pgfplotsretval
  \expandafter\xdef\csname RQc\RQy-\RQx\endcsname{\RQc}%
  \ifnum\RQx<7
    \expandafter\xdef\csname RQmax\RQy\endcsname{\fpeval{max(\csname RQmax\RQy\endcsname,\RQc)}}%
    \ifnum\fpeval{\RQc>0}=1
      \expandafter\xdef\csname RQkind\RQy\endcsname{\fpeval{\csname RQkind\RQy\endcsname+1}}%
    \fi
  \fi}%
\pgfplotsforeachungrouped \r in {0,...,4}{%
  \pgfplotsforeachungrouped \c in {0,...,7}{%
    \ifnum\c<7
      \edef\RQv{\csname RQc\r-\c\endcsname}%
      \ifnum\fpeval{\RQv>0}=1
        \ifnum\r=0
          \edef\RQi{\fpeval{round(25+60*(\RQv/\csname RQmax\r\endcsname)^0.5)}}%
        \else
          \edef\RQi{\fpeval{round(14+14*(\RQv/\csname RQmax\r\endcsname)^0.6)}}%
        \fi
      \else \def\RQi{0}\fi
    \else
      \ifnum\r=0
        \edef\RQi{\fpeval{round(25+60*(\csname RQkind\r\endcsname/7)^0.5)}}%
      \else
        \edef\RQi{\fpeval{round(14+14*(\csname RQkind\r\endcsname/7)^0.6)}}%
      \fi
    \fi
    \expandafter\xdef\csname RQi\r-\c\endcsname{\RQi}%
  }}%
\pgfplotsforeachungrouped \r/\col in {0/RQxcheck,1/RQneuri,2/RQgencog,3/RQhirgen,4/RQdeeprel}{%
  \expandafter\xdef\csname RQcol\r\endcsname{\col}}%
\def\RQfill#1#2{%
  \fill[fill=\csname RQcol#1\endcsname!\csname RQi#1-#2\endcsname!white, draw=white, line width=1.2pt]
    (axis cs:#2-0.5,#1-0.5) rectangle (axis cs:#2+0.5,#1+0.5);}%
\begin{figure}[t]
\centering
\begin{tikzpicture}
\def\RQp{\kern.3pt\raisebox{.15ex}{\tiny+}\kern.3pt}%
\def\RQcell{%
  \pgfmathtruncatemacro\RQr{\rowid}\pgfmathtruncatemacro\RQcc{\colid}%
  \edef\RQi{\csname RQi\RQr-\RQcc\endcsname}%
  \ifnum\RQr=0
    \ifnum\RQi>55 \color{white}\fi
    \bfseries
  \else
    \color{black!55}%
  \fi
  \ifnum\RQcc=7
    \csname RQkind\RQr\endcsname/7%
  \else\ifnum\RQi=0 \textcolor{black!40}{--}%
  \else
    \pgfmathprintnumber[fixed, precision=0, 1000 sep={\,}]{\cnt}%
  \fi\fi}
\begin{axis}[
  eval axis,
  width=0.98\linewidth, height=4.8cm,
  scale only axis=false,
  xmin=-0.5, xmax=7.5, ymin=-0.5, ymax=4.5,
  enlargelimits=false,
  clip=false,
  xtick={0,...,7},
  xticklabels={M, C, P, M\RQp C, M\RQp P, C\RQp P, M\RQp C\RQp P, {}},
  xticklabel pos=top,
  xticklabel style={font=\scriptsize, text height=1.5ex, text depth=.25ex, yshift=-2pt},
  ytick={0,...,4},
  yticklabels={\textbf{\tool}, \textcolor{black!60}{\neuri}, \textcolor{black!60}{\gencog}, \textcolor{black!60}{\hirgen}, \textcolor{black!60}{\deeprel}},
  yticklabel style={font=\footnotesize},
  y dir=reverse,
  tick style={draw=none},
  axis line style={draw=none},
  visualization depends on={\thisrow{count} \as \cnt},
  visualization depends on={\thisrow{y} \as \rowid},
  visualization depends on={\thisrow{x} \as \colid},
  nodes near coords={\RQcell},
  every node near coord/.append style={font=\footnotesize, anchor=center, text=black},
]
\pgfplotsinvokeforeach{0,...,4}{%
  \RQfill{#1}{0}\RQfill{#1}{1}\RQfill{#1}{2}\RQfill{#1}{3}%
  \RQfill{#1}{4}\RQfill{#1}{5}\RQfill{#1}{6}\RQfill{#1}{7}}
\addplot[only marks, mark=none,
         point meta=explicit symbolic]
  table[meta=count] {figures/data/rq1-constraint-layers.dat};
\draw[white, line width=5pt] (axis cs:6.5,-0.6) -- (axis cs:6.5,4.6);
\draw[RQxcheck!80!black, line width=.9pt, rounded corners=1.5pt]
  (axis cs:-0.487,-0.47) rectangle (axis cs:6.447,0.47)
  (axis cs:6.553,-0.47) rectangle (axis cs:7.487,0.47);
\draw[black!55, line width=.6pt]
  (axis cs:-0.45,-0.93) -- (axis cs:-0.45,-1.03) -- (axis cs:2.45,-1.03)
  node[midway, above=1pt, font=\footnotesize, text=black] {Single-Layer} -- (axis cs:2.45,-0.93);
\draw[black!55, line width=.6pt]
  (axis cs:2.55,-0.93) -- (axis cs:2.55,-1.03) -- (axis cs:6.45,-1.03)
  node[midway, above=1pt, font=\footnotesize, text=black] {Cross-Layer} -- (axis cs:6.45,-0.93);
\path (axis cs:7,-1.03) node[above=1pt, font=\footnotesize\bfseries, text=RQxcheck!80!black] {\vphantom{y}Kinds};
\end{axis}
\end{tikzpicture}
\caption{Constraints of \tool and the four baselines, grouped by the layers each constraint relates: model (M), compiler (C), and platform (P).
Shading compares the layer combinations within each approach, and {\it Kinds} counts the combinations that an approach covers.
\tool covers all seven combinations with 987 cross-layer constraints, whereas \neuri covers three and the other baselines cover only constraints over the model.}
\label{fig:rq1-constraint-matrix}
\end{figure}

\MyPara{Compilers under Test.}
We evaluate \tool on three DL compilers that accept ONNX models.
\begin{itemize}[leftmargin=*]
    \item TVM~\cite{tvm}.
    This compiler deploys models onto CPUs, GPUs, and specialized accelerators.
    We use TVM v0.23.0 at commit \codefont{75ce1b4}, which imports ONNX models into Relax or Relay and builds them for an LLVM target.
    It runs 73 distinct compilation passes.
    \item ONNX-MLIR~\cite{onnx-mlir}.
    This compiler lowers ONNX models through MLIR dialects onto CPUs and specialized AI accelerators.
    We use ONNX-MLIR v0.4.2 at commit \codefont{4a13c34}.
    It runs 32 distinct compilation passes.
    \item GeneSys~\cite{genesys}.
    This compiler maps ONNX models onto an accelerator with a systolic array and a SIMD unit.
    We use GeneSys at commit \codefont{8c5e488}.
    It runs 11 distinct compilation passes.
\end{itemize}

\MyPara{Baselines.}
We compare \tool with four existing model generators for DL compiler testing and \toolnoprioritization, a variant of \tool. Following prior work~\cite{nnsmith,neuri}, we run each approach for four hours on each compiler.
\begin{itemize}[leftmargin=*]
    \item \neuri~\cite{neuri}.
    This approach synthesizes PyTorch models with operator rules inferred from API invocations.
    We export its models to ONNX and compile them with the three compilers.
    \item \deeprel~\cite{deeprel}.
    This approach infers relational API pairs in PyTorch and fuzzes them.
    We export the generated programs to ONNX and compile them with the three compilers.
    \item \gencog~\cite{gencog}.
    This approach generates TVM computation graphs from constraint-based operator specifications.
    \item \hirgen~\cite{hirgen}.
    This approach generates high-level TVM IR programs by mutating hierarchical IRs.
    \item \toolnoprioritization. This variant of \tool disables the prioritization of constraints and generates models from all constraints uniformly at random.
\end{itemize}

\subsection{RQ1: Extracted Cross-Layer Constraints}
\label{sec:rq1}

\begin{table}[t]
\centering
\caption{Checks on the generation constraints and guards that \tool extracts.
The violability check separates the generation constraints that can be violated alone, which \tool targets for violation, from the constraints that the others imply, which \tool keeps for valid models.
Z3 checks 3,695 generation constraints and finds that 76.62\% of them can be violated alone. All 30 guards are consistent with their transformations.}
\label{tab:rq1-formal}
\small
\begin{tabular}{>{\raggedright\arraybackslash}p{2.6cm} p{7.6cm} r r}
\toprule
\textbf{Check} & \textbf{Description} & \textbf{\# Checked} & \textbf{Outcome} \\
\midrule
Independent violability & A generation constraint can be violated alone when a model violates it while the other constraints of its operator hold. Z3 searches for such an assignment. \tool targets only these constraints for violation and keeps the implied ones for valid models. & 3,695 & \makecell[tr]{2,831 violable\\864 implied} \\
\addlinespace[3pt]
Guard consistency & A guard is consistent when it holds on every result of a correct transformation, so its outcome describes what the compiler computes. Lean proves 5 arithmetic guards. Of the other guards, 8 compare a computed value directly with its requirement, and 17 restate the definition of their transformation, which we review. & 30 & 30 consistent \\
\bottomrule
\end{tabular}
\end{table}

We first evaluate which constraints \tool extracts and whether they are reliable enough to drive test generation and monitoring, since they decide which models \tool synthesizes and which intermediate states its guards check.
We compare the constraints that \tool extracts for TVM, ONNX-MLIR, and GeneSys with the constraints that the four baselines encode in their generators, and we group each constraint by the layers whose quantities it relates: the model, the compiler, and the platform.
We then run two checks on the extracted constraints: with the Z3 SMT solver~\cite{z3}, whether each generation constraint can be violated while the others hold, and with Z3, the Lean 4 theorem prover~\cite{lean4}, or manual review, whether each guard is consistent with the transformation it monitors.

Figure~\ref{fig:rq1-constraint-matrix} shows that \tool extracts constraints in all seven combinations of the model, compiler, and platform layers, whereas the baselines mainly encode constraints over the model.
Overall, \tool extracts 987 cross-layer constraints, which relate quantities from two or more layers.
For example, ONNX allows \codefont{ArgMax} to return the last position of the largest value when its attribute \codefont{select\_last\_index} is 1.
The TVM importer, however, raises \codefont{OpAttributeUnImplemented} for this setting.
From the importer, \tool therefore extracts the model--compiler constraint \codefont{select\_last\_index == 0}, so models for TVM keep the default value of this attribute.
Similarly, GeneSys compiles \codefont{Add} only when a codelet in its registry matches the ranks of the two input tensors $A$ and $B$.
\tool reads these codelets from the registry and extracts the model--compiler constraint \codefont{(rank(A), rank(B)) in \{(1,1), (2,2), (3,1), (4,4), ...\}}.
A model--compiler--platform constraint relates all three layers.
For \codefont{Add} on GeneSys, the constraint \codefont{TileSize * DataBits <= 523776} requires each tile of an input tensor to fit in the on-chip buffer \codefont{VMEM1}.
Here, the compiler selects the tile size for a dimension of the input tensor, the model gives the bit width of its data type, and the hardware specification gives the capacity of \codefont{VMEM1} as the product of the buffer depth, the accumulator width, and the number of SIMD lanes.
\tool extracts such constraints because its extraction rules also cover the compiler source and the hardware specification, in addition to the ONNX documentation.

In contrast, the 11 cross-layer rules of \neuri are written by hand to avoid known failures of specific tools.
For example, the input tensor of a reduction must have at least one dimension to avoid a known TVM crash, and padding must stay at most 255 to avoid an issue of the PyTorch exporter.
Such rules cover only the failures that have already been observed, and each new compiler or failure needs new rules written by hand.
\tool instead applies the same extraction rules to the documentation, source code, and hardware specification of each compiler, so its constraints scale to new compilers with little manual effort.
None of the baselines encodes a constraint that involves the platform.

A constraint that the other constraints imply cannot be violated alone, and a guard that fails on a correct compilation would record a behavior that the compiler does not exhibit, so we next run both checks, as Table~\ref{tab:rq1-formal} shows.
Z3 checks 3,695 generation constraints, and 76.62\% of them can be violated while all other constraints of the same operator hold.
For example, Z3 violates the TVM constraint \codefont{select\_last\_index == 0} of \codefont{ArgMax} with \codefont{select\_last\_index=4} on a rank-1 \codefont{float64} input tensor, and all other constraints of \codefont{ArgMax} hold for this assignment, so \tool can build a model that violates only this constraint.
The remaining 23.38\% are implied by others.
For example, \codefont{LeakyRelu} accepts \codefont{float16}, \codefont{float32}, and \codefont{float64} input tensors, but this constraint follows from the same type list for its output tensor together with the requirement that the input and output tensors have the same type.
\tool keeps such constraints for valid models, where they must still hold, and does not target them for violation.
All 30 guards pass this check.
Lean proves 5 arithmetic guards.
Of the other guards, 8 compare a computed value directly with its requirement, and we review the other 17, which restate the definition of their transformation.
For example, GeneSys pads a dimension size $n$ to the next multiple of an alignment $a$, \ie\ $p=\lceil n/a\rceil\cdot a$, and its guards require $p \bmod a = 0$, $p\geq n$, and $p-n<a$; Lean proves these three conditions for every natural number $n$ and every $a>0$, so these guards hold on every correct padding result.

\begin{figure}[t]
\centering
\pgfdeclarelayer{SKbg}\pgfsetlayers{SKbg,main}
\begin{tikzpicture}[font=\footnotesize, x=1cm, y=1cm]
\def\SKfracR{0.31}
\def\SKbar{0.07}
\def\SKgap{0.10}
\def\SKu{0.135}
\def\SKrow{0.025}
\def\SKcatgap{0.12}
\def\SKwC{1.7}
\def\SKwS{2.05}
\def\SKsplit{0.42}
\def\SKcgapmax{0.9}
\def\SKfontM{\scriptsize}
\def\SKfontR{\scriptsize}
\def\SKright{%
  \SKcat{catIV}{Input validation (8)}
  \SKcause{c0}{catIV}{1}{Operand type equality not checked}{, 1 fixed}
  \SKcause{c1}{catIV}{2}{Attribute domain not checked}{, 1 fixed}
  \SKcause{c2}{catIV}{2}{Operator arity not checked}{}
  \SKcause{c3}{catIV}{2}{Declared shape not checked}{}
  \SKcause{c4}{catIV}{1}{Index bounds not checked}{}
  \SKcat{catIL}{Incomplete lowering (6)}
  \SKcause{c5}{catIL}{1}{Float16 math function missing}{}
  \SKcause{c6}{catIL}{1}{Float16 infinity constant missing}{}
  \SKcause{c7}{catIL}{1}{BFloat16 constant missing}{}
  \SKcause{c8}{catIL}{1}{BFloat16 entry signature missing}{}
  \SKcause{c9}{catIL}{1}{Integer lowering missing}{}
  \SKcause{c10}{catIL}{1}{Legalization rule missing}{, 1 fixed}
  \SKcat{catDT}{Data-type handling (7)}
  \SKcause{c11}{catDT}{1}{8-bit integer exponent unhandled}{}
  \SKcause{c12}{catDT}{2}{Constant type fixed to float32}{}
  \SKcause{c13}{catDT}{3}{Unsigned integer mishandled}{}
  \SKcause{c14}{catDT}{1}{Integer input to float-only operation}{}
  \SKcat{catSI}{Shape and index computation (4)}
  \SKcause{c15}{catSI}{2}{Negative axis not normalized}{, 2 fixed}
  \SKcause{c16}{catSI}{2}{Per-axis attribute misapplied}{}
  \SKcat{catSC}{Schema conformance (9)}
  \SKcause{c17}{catSC}{1}{Multidirectional broadcasting mishandled}{, 1 fixed}
  \SKcause{c18}{catSC}{3}{Omitted attribute not defaulted}{, 1 fixed}
  \SKcause{c19}{catSC}{2}{Optional input or output mishandled}{}
  \SKcause{c20}{catSC}{2}{Axes input misread}{}
  \SKcause{c21}{catSC}{1}{Default attribute value rejected}{}
  \SKcat{catPI}{Pipeline infrastructure (2)}
  \SKcause{c22}{catPI}{1}{Graph copy fails on wrapped function}{}
  \SKcause{c23}{catPI}{1}{Empty memory region unhandled}{, 1 fixed}
}
\def\SKmid{%
  \SKsym{s0}{5}{Crash on invalid model}{}
  \SKsym{s1}{3}{Invalid model accepted}{}
  \SKsym{s2}{8}{Crash on valid model}{}
  \SKsym{s3}{7}{Valid model rejected}{}
  \SKsym{s4}{13}{Exception on valid model}{}
}
\def\SKleft{%
  \SKcomp{k0}{16}{ONNX-MLIR}{, 3 fixed}
  \SKcomp{k1}{9}{TVM}{, 4 fixed}
  \SKcomp{k2}{11}{GeneSys}{, 1 fixed}
}
\def\SKbands{%
  \SKfixed{catIV}{k0}{0}{s0}{0}{2}
  \SKflow{catIV}{k0}{2}{s0}{2}{3}
  \SKflow{catIV}{k0}{5}{s1}{0}{1}
  \SKflow{catIL}{k0}{6}{s2}{0}{5}
  \SKflow{catDT}{k0}{11}{s2}{5}{2}
  \SKfixed{catSI}{k0}{13}{s2}{7}{1}
  \SKflow{catDT}{k0}{14}{s3}{0}{2}
  \SKflow{catIV}{k1}{0}{s1}{1}{2}
  \SKfixed{catIL}{k1}{2}{s3}{2}{1}
  \SKflow{catDT}{k1}{3}{s3}{3}{3}
  \SKfixed{catSC}{k1}{6}{s3}{6}{1}
  \SKfixed{catSI}{k1}{7}{s4}{0}{1}
  \SKfixed{catSC}{k1}{8}{s4}{1}{1}
  \SKflow{catSI}{k2}{0}{s4}{2}{2}
  \SKflow{catSC}{k2}{2}{s4}{4}{7}
  \SKfixed{catPI}{k2}{9}{s4}{11}{1}
  \SKflow{catPI}{k2}{10}{s4}{12}{1}
  \SKfixed{catIV}{s0}{0}{c0}{0}{1}
  \SKfixed{catIV}{s0}{1}{c1}{0}{1}
  \SKflow{catIV}{s0}{2}{c1}{1}{1}
  \SKflow{catIV}{s0}{3}{c2}{0}{1}
  \SKflow{catIV}{s0}{4}{c3}{0}{1}
  \SKflow{catIV}{s1}{0}{c2}{1}{1}
  \SKflow{catIV}{s1}{1}{c3}{1}{1}
  \SKflow{catIV}{s1}{2}{c4}{0}{1}
  \SKflow{catIL}{s2}{0}{c5}{0}{1}
  \SKflow{catIL}{s2}{1}{c6}{0}{1}
  \SKflow{catIL}{s2}{2}{c7}{0}{1}
  \SKflow{catIL}{s2}{3}{c8}{0}{1}
  \SKflow{catIL}{s2}{4}{c9}{0}{1}
  \SKflow{catDT}{s2}{5}{c11}{0}{1}
  \SKflow{catDT}{s2}{6}{c12}{0}{1}
  \SKfixed{catSI}{s2}{7}{c15}{0}{1}
  \SKfixed{catIL}{s3}{0}{c10}{0}{1}
  \SKflow{catDT}{s3}{1}{c12}{1}{1}
  \SKflow{catDT}{s3}{2}{c13}{0}{3}
  \SKflow{catDT}{s3}{5}{c14}{0}{1}
  \SKfixed{catSC}{s3}{6}{c17}{0}{1}
  \SKfixed{catSI}{s4}{0}{c15}{1}{1}
  \SKflow{catSI}{s4}{1}{c16}{0}{2}
  \SKfixed{catSC}{s4}{3}{c18}{0}{1}
  \SKflow{catSC}{s4}{4}{c18}{1}{2}
  \SKflow{catSC}{s4}{6}{c19}{0}{2}
  \SKflow{catSC}{s4}{8}{c20}{0}{2}
  \SKflow{catSC}{s4}{10}{c21}{0}{1}
  \SKflow{catPI}{s4}{11}{c22}{0}{1}
  \SKfixed{catPI}{s4}{12}{c23}{0}{1}
}
\def\SKcm{28.45274}
\pgfmathsetmacro\SKW{\linewidth/\SKcm}
\def\SKnohyph{\hyphenpenalty=10000 \exhyphenpenalty=10000 }
\def\SKmeasure#1{\pgfinterruptpicture\global\setbox1=\hbox{#1}\endpgfinterruptpicture}
\def\SKrmax{0}\def\SKhmax{0}\def\SKnbug{0}\def\SKnsym{0}\def\SKncomp{0}
\def\SKcat#1#2{\SKmeasure{\SKfontR\bfseries #2}\pgfmathsetmacro\SKhmax{max(\SKhmax,\wd1/\SKcm)}}
\def\SKcause#1#2#3#4#5{\SKmeasure{\SKfontR #4 (#3#5)}\pgfmathsetmacro\SKrmax{max(\SKrmax,\wd1/\SKcm)}}
\def\SKsym#1#2#3#4{\pgfmathsetmacro\SKnbug{\SKnbug+#2}\pgfmathsetmacro\SKnsym{\SKnsym+1}}
\def\SKcomp#1#2#3#4{\pgfmathsetmacro\SKncomp{\SKncomp+1}}
\SKright\SKmid\SKleft
\pgfmathsetmacro\SKlabR{min(\SKrmax+0.15,\SKfracR*\SKW)}
\pgfmathsetmacro\SKxR{\SKW-\SKlabR-\SKgap-\SKbar}
\pgfmathsetmacro\SKxS{\SKwC+(\SKxR-\SKwC-\SKwS)*\SKsplit}
\pgfmathsetmacro\SKhx{min(\SKxR,\SKW-\SKhmax)}
\def\SKcur{0}\def\SKfirst{1}
\def\SKcat#1#2{%
  \pgfmathsetmacro\SKcur{\SKcur-\SKcatgap}%
  \node[anchor=north west, inner sep=0pt, outer sep=0pt, font=\SKfontR\bfseries,
        text=#1!75!black, fill=white, fill opacity=.85, text opacity=1] (SKh) at (\SKhx,\SKcur) {#2};
  \path (SKh.south); \pgfgetlastxy{\SKa}{\SKb}\pgfmathsetmacro\SKcur{\SKb/\SKcm-0.07}}
\def\SKcause#1#2#3#4#5{%
  \node[anchor=north west, inner sep=0pt, text width=\SKlabR cm, align=left,
        minimum height=#3*\SKu cm, font={\SKfontR\linespread{0.92}\selectfont\SKnohyph}]
    (#1) at ({\SKxR+\SKbar+\SKgap},\SKcur) {#4 \mbox{(#3#5)}};
  \path (#1.center); \pgfgetlastxy{\SKa}{\SKb}%
  \pgfmathsetmacro\SKt{\SKb/\SKcm+#3*\SKu/2}%
  \ifnum\SKfirst=1 \global\let\SKtop\SKt \gdef\SKfirst{0}\fi
  \fill[#2] (\SKxR,\SKt) rectangle ({\SKxR+\SKbar},{\SKt-#3*\SKu});
  \coordinate (#1-l) at (\SKxR,\SKt);
  \pgfmathsetmacro\SKbottom{\SKt-#3*\SKu}%
  \path (#1.south); \pgfgetlastxy{\SKa}{\SKb}\pgfmathsetmacro\SKcur{\SKb/\SKcm-\SKrow}}
\def\SKbox#1#2#3#4#5#6{%
  \pgfmathsetmacro\SKh{#4*\SKu}\pgfmathsetmacro\SKtw{#3-2*\SKbar-0.08}%
  \fill[black!9] (#2,\SKycur) rectangle ({#2+#3},{\SKycur-\SKh});
  \fill[black!50] (#2,\SKycur) rectangle ({#2+\SKbar},{\SKycur-\SKh});
  \fill[black!50] ({#2+#3-\SKbar},\SKycur) rectangle ({#2+#3},{\SKycur-\SKh});
  \coordinate (#1-l) at (#2,\SKycur);
  \coordinate (#1-r) at ({#2+#3},\SKycur);
  \node[align=center, text width=\SKtw cm, inner sep=0pt,
        font={\SKfontM\linespread{0.92}\selectfont\SKnohyph}]
    at ({#2+#3/2},{\SKycur-\SKh/2}) {\textbf{#5}\SKsep\mbox{(#4#6)}};
  \pgfmathsetmacro\SKycur{\SKycur-\SKh-\SKygap}}
\def\SKsym#1#2#3#4{\def\SKsep{\ }\SKbox{#1}{\SKxS}{\SKwS}{#2}{#3}{#4}}
\def\SKcomp#1#2#3#4{\def\SKsep{\\}\SKbox{#1}{0}{\SKwC}{#2}{#3}{#4}}
\SKright
\pgfmathsetmacro\SKygap{(\SKtop-\SKbottom-\SKnbug*\SKu)/(\SKnsym-1)}
\let\SKycur\SKtop
\SKmid
\pgfmathsetmacro\SKygap{min(\SKcgapmax,(\SKtop-\SKbottom-\SKnbug*\SKu)/(\SKncomp-1))}
\pgfmathsetmacro\SKycur{(\SKtop+\SKbottom)/2+(\SKnbug*\SKu+(\SKncomp-1)*\SKygap)/2}
\SKleft
\def\SKshape#1{\fill[#1]
    (\SKxA,\SKyA) .. controls (\SKxM,\SKyA) and (\SKxM,\SKyB) .. (\SKxB,\SKyB)
    -- (\SKxB,{\SKyB-\SKh}) .. controls (\SKxM,{\SKyB-\SKh}) and (\SKxM,{\SKyA-\SKh})
    .. (\SKxA,{\SKyA-\SKh}) -- cycle;}
\def\SKflow#1#2#3#4#5#6{%
  \path (#2-r); \pgfgetlastxy{\SKa}{\SKb}\pgfmathsetmacro\SKxA{\SKa/\SKcm}\pgfmathsetmacro\SKyA{\SKb/\SKcm-#3*\SKu}%
  \path (#4-l); \pgfgetlastxy{\SKa}{\SKb}\pgfmathsetmacro\SKxB{\SKa/\SKcm}\pgfmathsetmacro\SKyB{\SKb/\SKcm-#5*\SKu}%
  \pgfmathsetmacro\SKxM{(\SKxA+\SKxB)/2}\pgfmathsetmacro\SKh{#6*\SKu}%
  \begin{pgfonlayer}{SKbg}\SKshape{#1, fill opacity=.5, draw=white, line width=.3pt}\end{pgfonlayer}}
\def\SKfixed#1#2#3#4#5#6{\SKflow{#1}{#2}{#3}{#4}{#5}{#6}%
  \begin{pgfonlayer}{SKbg}\SKshape{pattern=north east lines, pattern color=#1!65!black}\end{pgfonlayer}}
\SKbands
\fill[black!50, fill opacity=.5] (0,\SKbottom) rectangle (0.4,{\SKbottom+\SKu});
\fill[pattern=north east lines, pattern color=black!70] (0,\SKbottom) rectangle (0.4,{\SKbottom+\SKu});
\node[anchor=west, inner sep=0pt, font=\SKfontM] at (0.5,{\SKbottom+\SKu/2}) {Fixed by the developers};
\end{tikzpicture}
\caption{Issues detected by \tool in each compiler, traced through their symptoms to their root causes. A valid model satisfies all of its constraints, and an invalid model, \ie a targeted-violation model, violates the prioritized constraint. Band widths are proportional to the number of issues, root causes are grouped and colored by category, and hatched bands mark the issues that the developers have since fixed. \tool detects 36 issues, and 72.22\% of them are missed by all baselines. Of these issues, 21 are compiler--platform interaction issues, and 85.71\% of them are missed by all baselines. The developers have since fixed eight issues.}
\label{fig:rq2-sankey}
\end{figure}

\rqsummary{1}{\tool extracts 987 cross-layer constraints, which relate two or more of the model, compiler, and platform layers. No baseline encodes a constraint that involves the platform. \gencog, \hirgen, and \deeprel encode no cross-layer constraint, and the 11 cross-layer rules of \neuri are written to avoid known failures and do not scale to new compilers. The checks determine that \tool has 2,831 generation constraints that can be violated alone and that all its 30 guards are consistent with the transformations that they monitor.}

\subsection{RQ2: Detected Issues and Root Causes}
\label{sec:rq2}

We next evaluate whether the tests generated by \tool detect issues in TVM, ONNX-MLIR, and GeneSys, and which of these issues the baselines miss.
We analyze the failing tests of the four-hour campaigns of \tool and the baselines on each compiler.
A failing test shows one of five symptoms.
On a valid model, which satisfies all of its constraints, the compiler may (1) crash on an assertion, an unreachable statement, or a segmentation fault, (2) raise an uncaught exception in its own code, such as an \codefont{IndexError}, or (3) reject the model with an error that it reports itself.
On a targeted-violation model, which violates the prioritized constraint, the expected behavior is to reject the model with a diagnostic.
Instead, the compiler may (4) crash or (5) compile the model without reporting the violation.
To count issues, we first group the failing tests that fail with the same error at the same code location.
For each group, we then locate the faulty code in the compiler source that causes the failure.
Different groups can lead to the same faulty code, so we count each piece of faulty code as one issue, however many tests and groups lead to it.
Failures where the compiler rejects an operator, a data type, or a rank with a diagnostic, or does not implement the operator at all, do not count as issues.
In contrast, when the earlier stages of the compiler accept a model but a later stage crashes on it, such as at an unreachable statement, the failure counts as an issue.

\begin{figure}[t]
\centering
\begin{tikzpicture}
\begin{groupplot}[
  group style={group size=3 by 1, horizontal sep=1.6cm},
  eval axis,
  width=0.29\linewidth, scale only axis=false,
  title style={font=\footnotesize\bfseries, text depth=0.25ex},
]
\nextgroupplot[title={TVM}, xbar stacked, height=4.6cm, bar width=6pt, xmin=0, xmax=100, xtick={0,25,50,75,100}, xlabel={Furthest stage (\% of tests)}, ymin=-0.6, ymax=5.6, ytick={0,1,2,3,4,5}, clip=false, axis x line*=bottom, axis y line*=left, xmajorgrids, y axis line style={draw=none}, ytick style={draw=none}, yticklabels={\hirgen,\gencog,\deeprel,\neuri,\toolnoprioritization,\tool}]
\addplot[fill=stageMV, draw=black!50, line width=0.3pt, forget plot] table[x=mv, y=pos, col sep=comma] {figures/data/rq3-stages-tvm.csv};
\addplot[fill=stageGT, draw=black!50, line width=0.3pt, forget plot] table[x=gt, y=pos, col sep=comma] {figures/data/rq3-stages-tvm.csv};
\addplot[fill=stageBL, draw=black!50, line width=0.3pt, forget plot] table[x=bl, y=pos, col sep=comma] {figures/data/rq3-stages-tvm.csv};
\addplot[fill=stageCG, draw=black!50, line width=0.3pt, forget plot] table[x=cg, y=pos, col sep=comma] {figures/data/rq3-stages-tvm.csv};
\node[anchor=west, font=\scriptsize, inner sep=1pt] at (axis cs:101,0) {42.6\%};
\node[anchor=west, font=\scriptsize, inner sep=1pt] at (axis cs:101,1) {84.4\%};
\node[anchor=west, font=\scriptsize, inner sep=1pt] at (axis cs:101,2) {95.0\%};
\node[anchor=west, font=\scriptsize, inner sep=1pt] at (axis cs:101,3) {73.9\%};
\node[anchor=west, font=\scriptsize, inner sep=1pt] at (axis cs:101,4) {88.1\%};
\node[anchor=west, font=\scriptsize, inner sep=1pt] at (axis cs:101,5) {96.3\%};
\nextgroupplot[title={ONNX-MLIR}, xbar stacked, height=4.6cm, bar width=6pt, xmin=0, xmax=100, xtick={0,25,50,75,100}, xlabel={Furthest stage (\% of tests)}, ymin=-0.6, ymax=5.6, ytick={0,1,2,3,4,5}, clip=false, axis x line*=bottom, axis y line*=left, xmajorgrids, y axis line style={draw=none}, ytick style={draw=none}, yticklabels={}, legend style={at={(0.5,-0.36)}, anchor=north, legend columns=4, /tikz/every even column/.append style={column sep=6pt}}, legend image code/.code={\draw[#1, draw=black!50, line width=0.3pt] (0cm,-0.08cm) rectangle (0.28cm,0.1cm);}]
\addplot[fill=stageMV, draw=black!50, line width=0.3pt] table[x=mv, y=pos, col sep=comma] {figures/data/rq3-stages-onnxmlir.csv};
\addlegendentry{Model validity check}
\addplot[fill=stageGT, draw=black!50, line width=0.3pt] table[x=gt, y=pos, col sep=comma] {figures/data/rq3-stages-onnxmlir.csv};
\addlegendentry{Graph transformation}
\addplot[fill=stageBL, draw=black!50, line width=0.3pt] table[x=bl, y=pos, col sep=comma] {figures/data/rq3-stages-onnxmlir.csv};
\addlegendentry{Backend lowering}
\addplot[fill=stageCG, draw=black!50, line width=0.3pt] table[x=cg, y=pos, col sep=comma] {figures/data/rq3-stages-onnxmlir.csv};
\addlegendentry{Code generation}
\node[font=\scriptsize, fill=white, inner sep=1.5pt] at (axis cs:50,0) {n/a};
\node[font=\scriptsize, fill=white, inner sep=1.5pt] at (axis cs:50,1) {n/a};
\node[anchor=west, font=\scriptsize, inner sep=1pt] at (axis cs:101,2) {99.8\%};
\node[anchor=west, font=\scriptsize, inner sep=1pt] at (axis cs:101,3) {96.0\%};
\node[anchor=west, font=\scriptsize, inner sep=1pt] at (axis cs:101,4) {80.7\%};
\node[anchor=west, font=\scriptsize, inner sep=1pt] at (axis cs:101,5) {89.2\%};
\nextgroupplot[title={GeneSys}, xbar stacked, height=4.6cm, bar width=6pt, xmin=0, xmax=100, xtick={0,25,50,75,100}, xlabel={Furthest stage (\% of tests)}, ymin=-0.6, ymax=5.6, ytick={0,1,2,3,4,5}, clip=false, axis x line*=bottom, axis y line*=left, xmajorgrids, y axis line style={draw=none}, ytick style={draw=none}, yticklabels={}]
\addplot[fill=stageMV, draw=black!50, line width=0.3pt, forget plot] table[x=mv, y=pos, col sep=comma] {figures/data/rq3-stages-genesys.csv};
\addplot[fill=stageGT, draw=black!50, line width=0.3pt, forget plot] table[x=gt, y=pos, col sep=comma] {figures/data/rq3-stages-genesys.csv};
\addplot[fill=stageBL, draw=black!50, line width=0.3pt, forget plot] table[x=bl, y=pos, col sep=comma] {figures/data/rq3-stages-genesys.csv};
\addplot[fill=stageCG, draw=black!50, line width=0.3pt, forget plot] table[x=cg, y=pos, col sep=comma] {figures/data/rq3-stages-genesys.csv};
\node[font=\scriptsize, fill=white, inner sep=1.5pt] at (axis cs:50,0) {n/a};
\node[font=\scriptsize, fill=white, inner sep=1.5pt] at (axis cs:50,1) {n/a};
\node[anchor=west, font=\scriptsize, inner sep=1pt] at (axis cs:101,2) {63.7\%};
\node[anchor=west, font=\scriptsize, inner sep=1pt] at (axis cs:101,3) {2.8\%};
\node[anchor=west, font=\scriptsize, inner sep=1pt] at (axis cs:101,4) {14.3\%};
\node[anchor=west, font=\scriptsize, inner sep=1pt] at (axis cs:101,5) {16.1\%};
\end{groupplot}
\end{tikzpicture}
\caption{Furthest compilation stage reached by the tests of each approach in the four-hour campaigns, as a percentage of all tests of the approach, with the share that reaches code generation at the end of each bar.
\tool outperforms \toolnoprioritization on all three compilers, \neuri on TVM and GeneSys, and \deeprel on TVM.
\neuri leads on ONNX-MLIR because it generates only valid models, and \deeprel leads on ONNX-MLIR and GeneSys because it generates only small models.
\gencog and \hirgen generate TVM IR directly, so they are inapplicable to ONNX-MLIR and GeneSys, and \tool also outperforms both on TVM.}
\label{fig:rq3-stage-reach}
\end{figure}

\MyPara{Symptoms.}
Figure~\ref{fig:rq2-sankey} traces the issues detected by \tool from the compiler that contains them through their symptoms to their root causes.
Overall, \tool detects 36 issues, and 72.22\% of them are not triggered by any test of the baselines.
Of these issues, 21 are compiler--platform interaction issues, which surface after the model validity check, and the baselines miss 85.71\% of them.
The 36 issues cover all five symptoms, whereas the issues that the baselines trigger cover only the three symptoms on valid models.
The issues that the baselines miss fall into two groups: 18 issues on valid models, and all eight issues on targeted-violation models, since the baselines never generate a model that breaks a constraint on purpose.

On valid models, the 18 issues that only \tool exposes occur on models that ONNX allows but the baselines never generate, such as models that omit an attribute or use a rare data type.
For example, \codefont{Transpose} reorders the axes of a tensor according to its \codefont{perm} attribute and reverses all axes when the model leaves \codefont{perm} out, so a $6\times3$ tensor becomes a $3\times6$ tensor (\ie\ \codefont{perm=[1,0]}).
When \tool generates such a model for GeneSys, the expected output has shape $3\times6$, but GeneSys raises an \codefont{IndexError} while importing the model.
Similarly, \codefont{Asinh} computes the inverse hyperbolic sine of each element, and ONNX allows it on \codefont{float16} tensors.
A model of \tool applies \codefont{Relu}, \codefont{Asinh}, and \codefont{Abs} to a \codefont{float16} tensor of shape $[8,1]$, and ONNX-MLIR accepts the model and lowers \codefont{Asinh}, but then aborts the process at an unreachable statement, although its earlier stages accept \codefont{float16}.
This is because the constraints that \tool extracts from the ONNX documentation record which attributes an operator may leave out and which data types it accepts.
Model synthesis samples from these domains, so it also generates models that rely on default attribute values or use \codefont{float16}, \codefont{bfloat16}, and unsigned integers.
In contrast, all 474 \codefont{Transpose} nodes in the GeneSys models of \neuri set \codefont{perm}, and all its \codefont{Conv} and \codefont{MaxPool} nodes set \codefont{strides}, so its models never rely on the default values of these attributes.
None of the models of \neuri and \deeprel uses \codefont{bfloat16}, which two ONNX-MLIR issues require, and \neuri never invokes \codefont{Asinh}.

On targeted-violation models, \tool exposes eight issues where a compiler skips a validity check and continues to compile a model that breaks a constraint.
Six of them lie in ONNX-MLIR, whose importer and verifiers accept models that break conditions that its lowering later relies on.
For example, \codefont{GatherElements} reads the elements of a tensor at the given indices, which must lie within the tensor.
\tool generates a TVM model that reads index 5 of a five-element tensor, whose valid indices are 0 to 4, and expects TVM to reject it, but TVM compiles the model, and the generated code reads memory outside the tensor.
Similarly, \codefont{GreaterOrEqual} compares two input tensors of the same data type, but a model of \tool compares a \codefont{float32} tensor with an \codefont{int64} tensor.
ONNX-MLIR accepts this model and then aborts on an assertion that both operands of the comparison have the same type.
This is because each targeted-violation model breaks the prioritized constraint and keeps all other applicable constraints, so the model is well formed except for the broken condition, and the compiler continues to the code that relies on it.

\MyPara{Root Causes.}
The 36 issues come from 24 root causes in six categories.
On valid models, schema conformance and data-type handling cause 16 issues, where a compiler mishandles an attribute or a data type that ONNX allows.
For example, when the model leaves \codefont{perm} out, GeneSys builds the reversed order of the axes from the sizes of the dimensions, $(3,6)$, instead of their positions, $(1,0)$ (\ie\ \codefont{reversed(shape)}), and then looks up dimension 3 of a two-dimensional tensor.
Shape and index computation causes four issues.
For example, \codefont{LpNormalization} normalizes a tensor along the axis given by its \codefont{axis} attribute, which defaults to the last axis and is written as $-1$, but ONNX-MLIR uses this value directly as an array index (\ie\ \codefont{reductionShape[axis] = 1}), so every model that relies on the default axis crashes the compiler.
On targeted-violation models, all eight issues come from missing input validation.
The crash of \codefont{float16} \codefont{Asinh} in ONNX-MLIR comes from incomplete lowering: an earlier stage of the compiler accepts an operator or a data type that a later stage does not implement.
ONNX-MLIR accepts \codefont{float16} \codefont{Asinh} and lowers it to its \codefont{Krnl} dialect, but the final lowering to LLVM only provides the \codefont{float32} and \codefont{float64} versions of the function.
\neuri misses these root causes because it never builds the models that reach them: it never invokes \codefont{LpNormalization} or \codefont{GatherElements}, and its exported models always set \codefont{perm} and \codefont{strides}.

The developers have since fixed eight issues, as marked in Figure~\ref{fig:rq2-sankey}.
For example, ONNX-MLIR now rejects a comparison such as \codefont{GreaterOrEqual} whose two operands have different data types.
It also adds the rank to a negative \codefont{axis} of \codefont{LpNormalization} before using it as an index, and TVM now normalizes a negative \codefont{axis} of \codefont{Hardmax}.

\begin{table}[t]
\centering
\caption{Categories of the 30 guards that \tool inserts into the padding, tiling, loop-unrolling, and operator-fusion sites of the three compilers.
Each guard example is the condition that the guard checks on an intermediate state.}
\label{tab:rq4-guards}
\small
\begin{tabular}{>{\raggedright\arraybackslash}p{2.5cm} >{\raggedright\arraybackslash}p{5.0cm} l r}
\toprule
\textbf{Guard Category} & \textbf{Description} & \textbf{Guard Example} & \textbf{\# Guards} \\
\midrule
Positive sizes & The alignment, tile size, and split factor that the compiler selects are positive. & \codefont{tileSize >= 1} & 4 \\
Aligned padding & The padded size is a multiple of the alignment that the target requires. & \codefont{p \% a == 0} & 2 \\
Padding covers the data & The padded size covers every operand and adds less than one alignment unit. & \codefont{p >= n and p - n < a} & 3 \\
Data fits the buffer & The tile kept in an on-chip buffer, including its padding, fits the capacity of the buffer. & \codefont{tile\_bits <= capacity\_bits} & 4 \\
Tiles cover the loop & The tiles visit every iteration of the original loop, and no tile is empty. & \codefont{(outer - 1) * inner < parent} & 5 \\
Unrolling keeps every iteration & The unrolled code holds one copy of the loop body per iteration, with the matching loop value. & \codefont{got.size() == k * extent} & 5 \\
Unrolling removes the loop variable & No statement of the unrolled code still refers to the loop variable. & \codefont{!UsesVar(result, loop\_var)} & 2 \\
Fusion keeps the interface & A fused function takes and returns values of the same types as the operators it replaces. & \codefont{equal(param\_type, arg\_type)} & 2 \\
Fusion leaves no dangling value & Every value used after fusion is still defined, and a fused function refers to no outside value. & \codefont{FreeVars(func).size() == 0} & 3 \\
\midrule
Total & & & 30 \\
\bottomrule
\end{tabular}
\end{table}

\begin{figure}[t]
\centering
\def\gdx{2.1}%
\def\gdX#1{\gdx+0.41*(#1)+0.3*(#1>8)+0.3*(#1>16)}%
\def\gdDot#1#2{\fill (#1,#2) circle (0.1);}%
\def\gdRing#1#2{\draw[line width=0.5pt] (#1,#2) circle (0.09);}%
\def\gdNone#1#2{\node[text=black!45] at (#1,#2) {--};}%
\def\gdrow#1#2#3#4{%
\pgfmathsetmacro{\xr}{\gdX{25}+0.25}%
\ifnum#4=1 \fill[black!9] (-0.1,#1-0.17) rectangle (\xr,#1+0.17); \fi
\node[anchor=west] at (0.15,#1) {#2};
\foreach \code [count=\ci from 0] in {#3}{%
\pgfmathsetmacro{\xx}{\gdX{\ci}}%
\if\code F\gdDot{\xx}{#1}\fi
\if\code O\gdRing{\xx}{#1}\fi
\if\code .\gdNone{\xx}{#1}\fi}}%
\def\gdgroup#1#2#3#4{%
\pgfmathsetmacro{\xL}{\gdX{#1}-0.15}\pgfmathsetmacro{\xR}{\gdX{#2}+0.15}\pgfmathsetmacro{\xM}{(\xL+\xR)/2}%
\node[font=\footnotesize\bfseries] at (\xM,#4+0.18) {#3};
\draw[line width=0.5pt] (\xL,#4)--(\xR,#4);}%
\begin{tikzpicture}[font=\footnotesize, gd legend/.style={anchor=west, text height=1.5ex, text depth=0.25ex}]
\pgfmathsetmacro{\xRule}{\gdX{25}+0.25}%
\foreach \h [count=\ci from 0] in {positive alignment,positive splits,aligned padding,covers source size,covers operand sizes,minimal padding,tile fits buffer,tiles fit memory,tiles cover loop,positive tile size,valid padding unit,aligned padding,block fits buffer,padding fits buffer,tiles cover loop,step matches tile,one copy per iteration,tiles cover loop,no empty tile,one copy per iteration,matching loop values,loop variable removed,same interface types,outputs remapped,no dangling value,no free variable}{%
\pgfmathsetmacro{\xx}{\gdX{\ci}}%
\node[rotate=70, anchor=base west, inner sep=1pt, font=\scriptsize] at (\xx+0.04,0.33) {\h};}
\node[anchor=west, font=\footnotesize\bfseries] at (0,0.42) {Approach};
\gdgroup{0}{8}{GeneSys}{2.55}
\gdgroup{9}{16}{ONNX-MLIR}{2.55}
\gdgroup{17}{25}{TVM}{2.55}
\draw[line width=0.9pt] (-0.1,2.9699999999999998)--(\xRule,2.9699999999999998);
\draw[line width=0.5pt] (-0.1,0.25)--(\xRule,0.25);
\node[anchor=west, font=\footnotesize\itshape] at (0,-0.06) {Each approach as published};
\gdrow{-0.40}{\tool}{F,F,F,F,F,F,F,F,F,F,F,F,F,F,F,F,F,O,O,O,O,O,F,F,F,O}{1}
\gdrow{-0.74}{\neuri}{O,O,O,O,O,O,O,O,O,O,O,O,O,O,O,O,O,O,O,O,O,O,O,O,O,O}{0}
\gdrow{-1.08}{\deeprel}{O,O,O,O,O,O,O,O,O,O,O,O,O,O,O,O,O,O,O,O,O,O,O,O,O,O}{0}
\gdrow{-1.42}{\gencog}{.,.,.,.,.,.,.,.,.,.,.,.,.,.,.,.,.,O,O,O,O,O,O,O,O,O}{0}
\gdrow{-1.76}{\hirgen}{.,.,.,.,.,.,.,.,.,.,.,.,.,.,.,.,.,O,O,O,O,O,O,O,O,O}{0}
\draw[line width=0.4pt, black!60] (-0.1,-2.04)--(\xRule,-2.04);
\node[anchor=west, font=\footnotesize\itshape] at (0,-2.35) {All approaches on the compilers with the guards of \tool};
\gdrow{-2.69}{\tool}{F,F,F,F,F,F,F,F,F,F,F,F,F,F,F,F,F,O,O,O,O,O,F,F,F,O}{1}
\gdrow{-3.03}{\neuri}{F,F,F,F,F,F,F,F,F,F,F,F,F,F,F,F,F,O,O,O,O,O,F,F,F,O}{0}
\gdrow{-3.37}{\deeprel}{F,F,F,F,F,F,F,F,F,O,O,O,O,O,O,O,O,O,O,O,O,O,F,F,F,O}{0}
\gdrow{-3.71}{\gencog}{.,.,.,.,.,.,.,.,.,.,.,.,.,.,.,.,.,F,F,F,F,F,F,O,O,F}{0}
\gdrow{-4.05}{\hirgen}{.,.,.,.,.,.,.,.,.,.,.,.,.,.,.,.,.,F,F,O,O,O,F,O,O,F}{0}
\draw[line width=0.9pt] (-0.1,-4.27)--(\xRule,-4.27);
\gdDot{0.2}{-4.62}\node[gd legend] at (0.4,-4.62) {evaluated by the tests};
\gdRing{4.03}{-4.62}\node[gd legend] at (4.23,-4.62) {not evaluated};
\gdNone{7.0}{-4.62}\node[gd legend] at (7.2,-4.62) {compiler not targeted by the approach};
\end{tikzpicture}
\caption{Guards evaluated by the tests of each approach in the four-hour campaigns.
The top block runs each approach as published, so only \tool inserts guards, and the bottom block runs all approaches on the same compilers with the guards of \tool. Guards that check the same condition share a column.
As published, only \tool observes intermediate compilation states, and its tests evaluate every guard of GeneSys and ONNX-MLIR and three of the four fusion guards of TVM.}
\label{fig:rq4-guards}
\end{figure}

\rqsummary{2}{\tool detects 36 issues across all five symptoms, and the baselines miss 72.22\% of them. Of these issues, 21 are compiler--platform interaction issues, and the baselines miss 85.71\% of them. The developers have since fixed eight issues.}

\subsection{RQ3: Exploration of Later Compilation Stages}
\label{sec:rq3}

Compiler--platform interaction issues surface in later compilation stages, where a compiler lowers operators to the target and allocates its memory and compute resources, and a test can only reveal such issues if it passes the earlier stages.
We therefore evaluate how far the tests of each approach go through the compiler and what share of them reaches the latest stage, code generation.
We run \tool, \toolnoprioritization, \neuri, and \deeprel for four hours on each compiler and divide each compiler into four stages: model validity check, graph transformation, backend lowering, and code generation.
For TVM, these stages correspond to the front end that reads the model, the graph-level passes that run before operators are lowered to loops, the lowering and the loop-level passes, and the final build.
For ONNX-MLIR, they correspond to the importer, the ONNX-dialect passes, the lowering to the \codefont{Krnl} dialect, and the conversion to LLVM.
For GeneSys, they correspond to the conversion from ONNX, the preprocessing passes, the codelet compilation stages, and code generation.
For each test submitted to a compiler, we record the furthest stage that the compilation enters, and we report the share of tests that end at each stage.
\gencog and \hirgen generate TVM IR programs directly, so they are inapplicable to ONNX-MLIR and GeneSys, and we run them for four hours on TVM only.

Figure~\ref{fig:rq3-stage-reach} shows that constraint prioritization raises the share of tests that reach code generation on all three compilers.
On TVM, 96.27\% of the tests of \tool reach code generation, while only 88.12\% of the tests of \toolnoprioritization do.
On ONNX-MLIR and GeneSys, the shares of \tool are 89.17\% and 16.12\%, while those of \toolnoprioritization are only 80.67\% and 14.28\%.
Both approaches build models from the same constraints, so this gain comes from prioritization alone.
For example, every TVM test that targets the output type of \codefont{Shrink} reaches code generation after completing 228 passes, so the shallowness of this constraint grows by only $1/228$ per test, and \tool prioritizes it 32 times, while \toolnoprioritization selects it only five times.
In contrast, the single test of \tool that violates the output count of \codefont{Neg} stops at the model validity check, so the shallowness of this constraint grows by 1, and \tool does not prioritize it again, while \toolnoprioritization selects it 18 times and ten of these tests stop at the model validity check.
This is because the shallowness of each constraint grows faster when its tests stop earlier, so \tool spends its time on constraints whose tests pass the earlier stages.
Prioritization also steers the targeted-violation models of \tool toward constraints that TVM does not check, so 87.02\% of these models reach code generation, while only 51.76\% of those of \toolnoprioritization do, and such models expose the missing validity checks reported in RQ2.

Compared with \neuri, \tool reaches code generation with a higher share on TVM and GeneSys, 96.27\% against 73.94\% and 16.12\% against 2.79\%.
This is because \tool builds each model from operators whose constraints it extracts from the compiler, so its GeneSys models only contain the 33 operator types that GeneSys supports.
In contrast, \neuri samples its operators from PyTorch, and its GeneSys models contain 95 operator types.
For example, 61.60\% of the GeneSys tests of \neuri stop at the model validity check because they contain an operator that GeneSys does not support, such as \codefont{Cast}, \codefont{Sin}, or \codefont{Round}.
\neuri reaches a higher share only on ONNX-MLIR, 96.01\% against 89.17\%.
This is because \neuri generates only valid models, while 12.75\% of the tests of \tool use targeted-violation models, which the compiler is expected to reject early.
On valid models, the tests of \tool reach code generation in 96.82\% of cases, on par with \neuri.

Compared with \deeprel, \tool reaches code generation with a higher share on TVM, 96.27\% against 94.95\%.
\deeprel reaches higher shares on ONNX-MLIR and GeneSys, 99.82\% and 63.68\%, because it generates only small models.
At least 92\% of its tests that reach code generation apply a single \codefont{Relu} or a \codefont{Clip} followed by \codefont{Relu}.
These tests cover at most five operator types, against 105 and 114 for \tool on TVM and ONNX-MLIR.
On GeneSys, all tests of \deeprel that reach code generation then fail there, because none of their models has weights.

On TVM, \tool also reaches code generation with a higher share than \gencog and \hirgen, 96.27\% against 84.35\% and 42.65\%.
Most of their other tests stop at graph transformation, 13.65\% for \gencog and 57.16\% for \hirgen.
For example, 54.57\% of the tests of \hirgen stop during constant folding, and 13.56\% of the tests of \gencog stop when TVM changes the data layout of operators.

On GeneSys, 82.80\% of the tests of \tool stop at the model validity check.
One reason is an issue that \tool itself detects: GeneSys fails to copy a graph that contains an \codefont{Identity} node, which \tool adds to extend the graph around the target operator.
This failure stops 34.90\% of the GeneSys tests of \tool.
Overall, 52.65\% of the GeneSys models of \tool contain such a node, and none of their tests reaches code generation, while the other tests of \tool reach code generation in 34.05\% of cases.
Feeding a detected issue back into model synthesis, so that later models no longer contain the node that triggers it, would let these tests reach the later stages as well.

\rqsummary{3}{Constraint prioritization raises the share of tests that reach code generation over \toolnoprioritization by 8.15, 8.50, and 1.84 percentage points on TVM, ONNX-MLIR, and GeneSys. This is because the shallowness of each constraint grows faster when its tests stop earlier, so \tool spends its time on constraints whose tests pass the earlier stages.}

\subsection{RQ4: Effectiveness of Guard-Based Monitoring}
\label{sec:rq4}

Guards record how the intermediate states of a compilation relate to the monitoring constraints, which the compilation result alone does not reveal.
A guard only observes a state when a test drives the compiler through the transformation that the guard instruments.
We therefore evaluate which guards the tests of each approach evaluate in the four-hour campaigns, and a guard counts as evaluated when at least one test evaluates it.

Table~\ref{tab:rq4-guards} reports the result of guard insertion.
\tool inserts 30 guards into the padding, tiling, loop-unrolling, and operator-fusion code of the three compilers, and we group them into nine categories by the condition they check.
These categories instantiate the monitoring categories of Table~\ref{tab:constr-category} at the sites where the compilers compute the values.
For padding, two {\it aligned padding} guards require the padded size to be a multiple of the alignment that the target requires, which instantiates hardware alignment.
Three {\it padding covers the data} guards require the padded size to cover the original size and to add less than one alignment unit.
For tiling, five {\it tiles cover the loop} guards require the tiles to visit every iteration of the original loop, which instantiates transformation results.
Four {\it data fits the buffer} guards require each tile, including its padding, to fit the on-chip buffer that holds it, which instantiates resource capacity.
Another four {\it positive sizes} guards require the alignment, tile size, and split factor that the compiler selects to be positive.
For loop unrolling, five guards require the unrolled code to hold one copy of the loop body per iteration, and two require that no statement still refers to the loop variable.
For operator fusion, two guards require a fused function to keep the types of the operators it replaces, and three require that it refers to no value outside itself.
Each guard outcome therefore names the transformation and the condition that it records, such as whether a tile fits its buffer.

Figure~\ref{fig:rq4-guards} shows that the baselines, as published, observe no intermediate state, because they do not insert any guard into the compilers.
In contrast, the tests of \tool evaluate every guard of GeneSys and ONNX-MLIR, and three of the four fusion guards of TVM.
Two tests show the same guard behavior when their compilations evaluate the same guards on the same codelets, operators, or fused blocks, with the same outcomes.
Under this definition, the tests of \tool show 269 distinct behaviors: 235 on GeneSys, 31 on TVM, and three on ONNX-MLIR.
In contrast, the compilation result only tells whether each of these tests fails and at which stage.

To compare the tests rather than the instrumentation, we also run all approaches on the compilers with the guards of \tool.
The guards that the tests reach then depend on which transformations their models drive the compiler through.
On ONNX-MLIR, the tests of \tool and \neuri evaluate all eight guards, while the tests of \deeprel evaluate none.
This is because the models of \deeprel only apply \codefont{Relu}, \codefont{Clip}, and pooling operators, whose lowering does not go through tiling, padding, or unrolling.
On TVM, the tests of \tool, \neuri, and \deeprel evaluate the fusion guards, and the tests of \gencog and \hirgen, which generate TVM IR directly, also evaluate the tiling guards. The tests of \gencog also evaluate the unrolling guards.
On GeneSys, the tests of \tool, \neuri, and \deeprel all evaluate every guard, but the tests of \tool reach them through more parts of the compiler.
For example, they evaluate the padding guards in 40 codelets, against 18 for \neuri and four for \deeprel, including convolution, matrix multiplication, and softmax codelets that no baseline reaches.
This is because \tool builds each model around the operator of the prioritized constraint and uses only operators that GeneSys supports, so its models reach the codelets of operators such as \codefont{Conv}, \codefont{Gemm}, and \codefont{MatMul}.

\begin{table*}[t]
\centering
\small
\caption{Effort to extend \tool to a new compiler. Doc Share: share of the constraints of the new compiler that come from the ONNX documentation alone and are reused without new work; Config: lines of configuration newly written for the new compiler, which list its source files, functions, and guard sites; Code: code newly written for the new compiler, as a share of all code of \tool.}
\label{tab:extensibility-effort}
\begin{tabular}{lrrrrrr}
\toprule
\multirow{2}{*}{\textbf{Extension}} & \textbf{Reused} & \multicolumn{3}{c}{\textbf{Constraint Extraction}} & \multicolumn{2}{c}{\textbf{Guard Insertion}} \\
\cmidrule(lr){2-2}\cmidrule(lr){3-5}\cmidrule(lr){6-7}
 & \textbf{Doc Share} & \textbf{Config} & \textbf{Code} & \textbf{AST Parsing} & \textbf{Config} & \textbf{Code} \\
\midrule
GeneSys $\rightarrow$ ONNX-MLIR & 79.43\% & 50 LOC & 8.84\% & 0.63 s & 70 LOC & 1.56\% \\
ONNX-MLIR $\rightarrow$ TVM & 98.74\% & 12 LOC & 3.93\% & 1.98 s & 48 LOC & 1.04\% \\
\bottomrule
\end{tabular}
\end{table*}

\rqsummary{4}{Guards reveal intermediate compilation states that coarse-grained signals miss: the tests of \tool show 269 distinct guard behaviors across the three compilers, whereas a compilation result only tells whether a test fails and at which stage. As published, the baselines observe no such state.}

\section{Discussion}
\label{sec:threats}
\subsection{Extensibility}
\label{sec:rq5}
\label{sec:discussion-extensibility}

Extending \tool to a new compiler takes little effort, because most of its constraints and all of its testing components carry over.
A compiler that accepts ONNX models reuses the constraints extracted from the ONNX documentation.
As Table~\ref{tab:extensibility-effort} shows, these constraints make up 79.43\% of the constraints for ONNX-MLIR and 98.74\% for TVM.
Model synthesis, constraint prioritization, and the Z3 and Lean checks are also reused unchanged.

Only the constraints from the compiler source and the platform need new work.
For these constraints, the user writes two short configurations of 12 to 70 lines each, which name the source files, functions, and guard sites of the compiler, and provides the parameters of the platform, such as memory capacity, data width, and array dimensions.
The new code parses the structures specific to the compiler, such as the importer of ONNX-MLIR and the frontend of TVM, and inserts the guards.
This code makes up only 10.40\% of the code of \tool for ONNX-MLIR and 4.97\% for TVM.
Parsing the source of each compiler into ASTs with \codefont{tree-sitter}~\cite{treesitter} takes only 0.63 seconds for ONNX-MLIR and 1.98 seconds for TVM.

Adding constraints for a compiler that \tool already tests only needs new extraction rules.
For example, a new documentation rule can match an operator attribute that the current rules miss.
A new AST rule can cover a transformation that no guard checks yet, such as vectorization.
A new hardware specification adds the resource and alignment limits of another platform.
Each new constraint records its predicate, its source, and the conditions under which it applies, such as the lowering path and the target.
It also maps each of its variables to a model attribute, a compiler value, or a platform parameter.
\tool then checks the new constraints in the same way as the extracted ones: each generation constraint for independent violability and each guard for consistency with its transformation.

\subsection{Relationship to LLM-Based DL System Testing}
\label{sec:discussion-llm}

\tool is complementary to LLM-based testing for DL systems.
TitanFuzz~\cite{titanfuzz} and FuzzGPT~\cite{fuzzgpt} prompt LLMs to generate programs that call the APIs of DL libraries.
WhiteFox~\cite{whitefox} asks an LLM to summarize, from the source code of a compiler optimization, which inputs trigger the optimization, and then to generate programs with such inputs.
These approaches rely on the LLM to infer the constraints on inputs from code and text, and they detect issues only from the final result of a program, such as a crash or a different output under two configurations.
In contrast, \tool extracts each constraint deterministically from the source that states it, records whether each test satisfies or violates the constraint, and observes intermediate compilation states with guards.

\tool and LLM-based testing can be combined in both directions.
First, an LLM can infer constraints that the extraction rules of \tool do not match, such as constraints stated only in the natural-language summary of an operator or in the comments of compiler code.
\tool can then check these constraints in the same way as the extracted ones before model synthesis uses them.
Second, an LLM-based generator can generate programs that satisfy or violate the constraints of \tool and compile these programs with the guards of \tool.
Its tests then also record the intermediate compilation states that the final result does not reveal.

\subsection{Threats to Validity}
\label{sec:threats-validity}

\MyPara{Internal Validity.}
The validity of the generated tests depends on the correctness of the extracted constraints and guards.
An incorrect generation constraint would mislabel a test as valid or as a targeted violation, and an inconsistent guard would record a behavior that a correct compilation does not have.
To reduce this threat, each constraint records the documentation entry, source location, or hardware specification field that it comes from, and we check the constraints and guards before the campaigns.
\tool also evaluates the constraints again on each saved model, so a requested violation that the model does not realize is not counted as one.
The setup of the baselines can also affect the comparison.
We run each baseline with its published generation method and convert the PyTorch programs of \neuri and \deeprel to ONNX.
Our only change to a baseline restores two commented-out lines of \deeprel that save the programs it runs successfully, so that these programs are also compiled.

\MyPara{Construct Validity.}
We count issues by faulty code location, but one location can hold more than one defect, and one defect can surface at more than one location.
We therefore also analyze the root cause of each issue.
Whether a compiler failure is an issue or an expected rejection can be ambiguous.
When the compiler rejects a model with an error message that explains why, such as an unsupported data type, we treat the failure as expected behavior rather than an issue.
In contrast, when the earlier stages of the compiler accept a model but a later stage crashes on it, we count the failure as an issue.
The stage that a test reaches depends on how we divide each compiler into stages.
We place the boundaries between the parts that each compiler already separates and use the same boundaries for all approaches.

\MyPara{External Validity.}
Our evaluation covers TVM, ONNX-MLIR, and GeneSys, which differ in their compiler stacks and targets, and the quantitative results may differ for other compilers and platforms.
Existing constraints of \tool come from the ONNX documentation, the source of the three compilers, and hardware specifications, and its AST extraction rules cover four common transformations.
New extraction rules for documentation, compiler code, and hardware specifications extend this constraint set.

\MyPara{Conclusion Validity.}
Following prior work~\cite{nnsmith,neuri}, we run each approach once for four hours on each compiler, so another run may reach different operators and expose different issues.
We compare \tool with \toolnoprioritization, which shares every setting except constraint prioritization, and report the gains that hold on all three compilers.

\section{Related Work}
\label{sec:related-work}
\MyPara{Constraint-Based Model Generation for Testing DL Compilers.}
Most existing techniques test DL compilers by generating models that satisfy operator-level constraints, so that the generated models are type-valid by construction.
The main difference among the approaches in this category lies in where the constraints are derived from~\cite{deeprel,nnsmith,neuri,docter,titanfuzz,fuzzgpt,gencog,whitefox,oatest}.
They have proven effective at generating valid models that cover a wide range of operators and shapes.
For example, NeuRI~\cite{neuri} infers operator constraints from execution traces and applies concolic solving to generate models that satisfy the inferred relations.
GenCoG~\cite{gencog} instead introduces a DSL for developers to explicitly specify operator constraints and incrementally solves them to construct valid models.
In all of them, constraints mainly shape the input space, and bugs are detected through pass/fail signals or differential mismatches.
In contrast, \tool also extracts constraints from the compiler implementation and the hardware specification, and it uses them beyond shaping the input space.
It synthesizes models that satisfy all applicable constraints or violate a targeted one, prioritizes the constraints whose tests reach later compilation stages, and inserts guards into compilation passes to record how intermediate compilation states relate to quantities of the compiler and the platform.

\MyPara{Transformation-Based Differential Testing of DL Compilers.}
Transformation-based techniques derive new tests by mutating or reconstructing existing programs or models~\cite{cradle,audee,lemon,muffin,luo2021graph,polyjuice,hirgen,modelmeta}.
The applied rules are equivalence-preserving, so the original and the transformed version are expected to agree, and a disagreement is reported as a bug.
These techniques differ mainly in the level at which the transformation is applied.
For example, PolyJuice~\cite{polyjuice} applies arithmetic and structural rewrite rules to construct semantically equivalent computation graphs.
HirGen~\cite{hirgen} applies function rewrites to add wrappers in IRs, and ModelMeta~\cite{modelmeta} generates models that have consistent outputs but different model structures and calculation logic.
The same idea also applies to the transformations that the compiler itself performs.
DiTOX~\cite{ditox} runs the passes of the ONNX Optimizer on real models and traces an output difference back to a single pass, which is possible because these passes are graph-based and do not alter numeric precision.
Apart from crashes and rejections, these techniques report a bug only when two runs disagree, so a compilation that is wrong but still produces a normal output, such as a model that violates a constraint but is accepted by the compiler, stays invisible to them.
In contrast, \tool does not compare two runs.
Its targeted-violation models reveal when a compiler accepts a model that breaks a constraint, and its guards record what each monitored pass computes, independently of the final output.

\MyPara{Test Oracles for DL Compilers.}
Beyond generating tests, a testing technique needs an oracle to decide whether a compilation is correct.
Crashes and output mismatches provide a readily available oracle, but they miss failures that leave the output intact.
Existing techniques therefore obtain a finer oracle from outside the compiler, and they differ in where it comes from.
TAIDL~\cite{taidl} specifies the ISA of a tensor accelerator in a dedicated language and automatically generates a functional simulator from that specification, which serves as a reference for code compiled to the accelerator.
Triton-Sanitizer~\cite{tritonsanitizer} checks the generated kernel instead.
It builds symbolic expressions for memory addresses and masks, verifies them with an SMT solver, and reports memory access errors together with the symbolic computation that produced the faulty address.
Such oracles are precise and expose faults that crash-based signals miss.
Each of them, however, operates at one level of the stack only, namely the accelerator ISA or the emitted kernel.
\tool instead records finer-grained signals inside the compiler.
It extracts constraints that are already stated in the model documentation, the hardware specification, and the compiler source, and inserts guards into compilation passes.
This also changes what can be observed: a kernel-level sanitizer reports a memory access error after code generation, whereas a guard of \tool, such as one that relates a tile computed by the compiler to the capacity of an on-chip buffer, records this relation at the pass that computes the tile.

\MyPara{Specification and LLM-Based Test Generation.}
The constraints used for testing must themselves be obtained, and prior work differs in how that is done.
They are written by hand, parsed from natural-language documents, inferred from executions, or produced by an LLM.
TAIDL~\cite{taidl} asks architects to write an ISA specification, and DESIL~\cite{desil} defines undefined-behavior elimination rules by manually reviewing the MLIR documentation for each supported operation.
DocTer~\cite{docter} constructs rules over the dependency parse trees of DL API descriptions and extracts input constraints for API functions across TensorFlow, PyTorch, and MXNet, while NeuRI~\cite{neuri} derives them from execution traces rather than from a written specification.
TitanFuzz~\cite{titanfuzz} and FuzzGPT~\cite{fuzzgpt} rely on the learned priors of LLMs in place of an explicit specification to generate programs for DL library testing, and WhiteFox~\cite{whitefox} analyzes the source code of compiler optimizations with an LLM and summarizes the conditions under which each optimization is triggered.
Automating this step removes most of the manual effort and scales to large operator sets.
What the obtained constraints describe, however, remains local, as each one states what a single artifact accepts or computes, such as one API, one operator, or one optimization.
\tool extracts constraints that span model semantics, compilation transformations, and hardware feasibility, and cross-references them, so that a single constraint can relate quantities from different layers.
No single specification relates a value computed inside the compiler to a limit declared in the hardware specification, and these are the relations that compiler--platform interaction issues violate.

\MyPara{Compiler Bug Deduplication.}
Different failing tests of a testing campaign can expose the same compiler defect, so they must be grouped before developers investigate them.
BugLens~\cite{buglens2026} uses the failure-inducing commits found by bisection, together with the optimizations that trigger each bug, to select representative failing tests for investigation.
This motivates grouping failing tests by evidence beyond their observed symptoms, since two tests with the same symptom do not necessarily expose the same defect.
We follow the same principle in our evaluation: we group failing tests by the faulty code location in the compiler source rather than by their error messages, and count each faulty code location once as an issue.

\MyPara{Constraint-Based General-Purpose Testing.}
Constraints are also used for testing general-purpose compilers and language processors, where a test program must satisfy the grammar as well as semantic conditions that the grammar cannot express.
These approaches differ in how those semantic conditions reach the tool.
For example, ISLa~\cite{isla} produces inputs satisfying logical constraints expressed over a grammar, and it relies on an SMT solver to enforce both syntactic and context-sensitive semantic constraints.
Its companion tool, ISLearn~\cite{islearn}, infers such constraints automatically, but requires users to provide domain-specific constraint schemata in a dedicated DSL.
Similarly, Fandango~\cite{fandango} adopts a constraint-guided generation paradigm, but uses Python as a host language for manually specifying user-defined constraints.
These approaches demonstrate the effectiveness of constraint-aware input generation, and they remain independent of any particular compiler.
They also depend on either hand-written DSL specifications or predefined templates to infer constraints.
\tool targets a domain in which such specifications already exist.
It extracts constraints from documentation and the codebase, which reduces the manual effort and lowers the barrier to deploying constraint-guided testing on large, real-world compilers.

\section{Conclusion}
\label{sec:conclusion}
We presented \tool, a DL compiler testing approach that reveals compiler--platform interaction issues by using cross-layer constraints over the model, the compiler, and the platform both to guide test generation and to monitor compilation.
\tool deterministically extracts these constraints from documentation, compiler code, and hardware specifications, checks that each constraint can be violated on its own and that each guard is consistent with the transformation it monitors, and then synthesizes models that satisfy all applicable constraints or violate a targeted one.
Compilation feedback steers the prioritization of constraints toward later compilation stages, and guards inserted into the compilers record how intermediate states relate to quantities of the compiler and the platform, which distinguishes compilations that coverage and compilation outcomes treat as the same.
On TVM, ONNX-MLIR, and GeneSys, \tool detects 36 issues, including 21 compiler--platform interaction issues.
The four baselines miss 85.71\% of these interaction issues.
Constraint prioritization raises the share of tests that reach code generation by up to 8.50 percentage points.
The guards of \tool observe 269 distinct behaviors of intermediate compilation states.

\bibliographystyle{ACM-Reference-Format}
\bibliography{main}

\end{document}